\documentclass[aps,prx,reprint,groupedaddress,superscriptaddress,floatfix,longbibliography]{revtex4-2}

\usepackage[T1]{fontenc}
\usepackage{newtxtext}
\usepackage{newtxmath}

\usepackage{amsmath}
\usepackage{mathtools}
\usepackage{amsfonts}
\usepackage{bbold}

\usepackage{color}
\usepackage{xcolor}
\usepackage[colorlinks,citecolor=blue,linkcolor=blue,urlcolor=blue]{hyperref}

\usepackage{graphicx}
\usepackage[all]{hypcap}

\usepackage{enumitem}

\usepackage{tabularx}
\usepackage{makecell}

\begin{document}

\title{Optimization via Quantum Preconditioning}

\author{Maxime Dupont}
\email[Corresponding author:~]{mdupont@rigetti.com}
\affiliation{Rigetti Computing, 775 Heinz Avenue, Berkeley, California 94710, USA}

\author{Tina Oberoi}
\email[Current affiliation:~University of Chicago, Chicago, Illinois, USA]{}
\affiliation{Rigetti Computing, 775 Heinz Avenue, Berkeley, California 94710, USA}

\author{Bhuvanesh Sundar}
\affiliation{Rigetti Computing, 775 Heinz Avenue, Berkeley, California 94710, USA}

\begin{abstract}
    State-of-the-art classical optimization solvers set a high bar for quantum computers to deliver utility in this domain. Here, we introduce a quantum preconditioning approach based on the quantum approximate optimization algorithm. It transforms the input problem into a more suitable form for a solver with the level of preconditioning determined by the depth of the quantum circuit. We demonstrate that best-in-class classical heuristics such as simulated annealing and the Burer-Monteiro algorithm can converge more rapidly when given quantum preconditioned input for various problems, including Sherrington-Kirkpatrick spin glasses, random $3$-regular graph maximum-cut problems, and a real-world grid energy problem. Accounting for the additional time taken for preconditioning, the benefit offered by shallow circuits translates into a practical quantum-inspired advantage for random $3$-regular graph maximum-cut problems through quantum circuit emulations. We investigate why quantum preconditioning makes the problem easier and test an experimental implementation on a superconducting device. We identify challenges and discuss the prospects for a hardware-based quantum advantage in optimization via quantum preconditioning.
\end{abstract}

\maketitle

\section{Introduction}

Optimization problems over discrete variables are ubiquitous~\cite{papadimitriou1998,Korte2012,Kochenberger2014}; however, they continue to pose a significant challenge for traditional computing paradigms. The absence of provably efficient algorithms in the general case unless $\textrm{P}=\textrm{NP}$~\cite{10.5555/1074100.1074233,Barahona1982} makes heuristic solvers the practical method of choice. Celebrated examples include physics-inspired approaches, such as simulated annealing (SA)~\cite{kirkpatrick1983optimization} or parallel tempering~\cite{hukushima1996exchange}, as well as heuristics that build on approximate methods with performance guarantees, like the Burer-Monteiro (BM) solver~\cite{burer2002rank,Burer2003} based on the Goemans-Williamson (GW) algorithm for the maximum-cut problem~\cite{Goemans1995,williamson2011}. These heuristics represent the state of the art in discrete optimization on a range of problems~\cite{DunningEtAl2018}.

The advent of quantum technologies as a new computing paradigm~\cite{AndrewSteane_1998,Ladd2010} has sparked interest in exploring whether discrete optimization can leverage quantum phenomena for improved efficiency~\cite{Abbas2024}. On the one hand, quantum adiabaticity~\cite{Farhi2000,Farhi2001} serves as the foundation for quantum annealers and, more broadly, analog quantum computation~\cite{RevModPhys.80.1061,Hauke_2020}. Analog platforms, through empirical observation, have demonstrated an asymptotic scaling advantage over classical methods such as SA and parallel tempering on specific classes of problems~\cite{PhysRevX.8.031016,Ebadi2022,King2023,Bauza2024}; however, an absolute advantage in computational time has not yet been realized at currently addressed problem sizes on current quantum hardware. On the other hand, digital quantum computers operating via programmable quantum logical operations typically leverage the variational principle to address discrete optimization~\cite{Cerezo2021}. In that context, a relevant algorithm is the quantum approximate optimization algorithm (QAOA)~\cite{Farhi2014,BLEKOS20241}. In certain limiting cases and for specific problem types, the QAOA can provide performance guarantees on the accuracy of the returned solution~\cite{Farhi2014,Farhi2022,basso_et_al:LIPIcs.TQC.2022.7,PhysRevA.103.042612,Montanaro2024}. Additionally, the QAOA can be used as a heuristic, where classical emulations support evidence of a scaling advantage for certain problems that is yet to be realized in practice~\cite{shaydulin2024evidence,Barrera2024}. In addition to the QAOA, other quantum algorithms with a potential scaling advantage have been advanced~\cite{Hastings2018shortpathquantum,10.1145/3564246.3585203,Kapit2023,Jordan2024}. While the evidence for a scaling advantage is promising, transforming this into a tangible advantage for quantum computers to deliver utility remains a substantial engineering challenge. Alongside hardware development, further progress in quantum algorithms is equally important in this endeavor.

Here, we introduce the idea of quantum preconditioning for discrete optimization problems (Sec.~\ref{sec:quantum_precond}). In the classical context, preconditioning seeks to transform a problem into a more suitable form for a solver. It aims to make the preconditioned problem easier to solve than its original version by, e.g., converging faster to a solution. A notable example arises in state-of-the-art classical linear system solvers, where classical preconditioning is instrumental to their performance~\cite{Turkel1999,Wathen_2015}. A related idea for solving mixed integer programming problems is ``presolving''~\cite{Andersen1995,Gamrath2015}: this involves preprocessing techniques being applied to the input problem to make it more tractable for a solver.

We propose a quantum preconditioner for quadratic unconstrained binary optimization problems~\cite{lucas2014} based on the QAOA with $p$ layers which works as follows. The goal is to extremize a quadratic objective function $C(\boldsymbol{z})$ over $N$ discrete variables $\boldsymbol{z}=(z_1,z_2,\ldots,z_N)^T$, where the objective function is encoded by an adjacency matrix $\mathsf{W}\in\mathbb{R}^{N\times N}$. First, we execute the QAOA on the original $N$-variable problem encoded by $\mathsf{W}$. Then, we use the resulting quantum state to estimate two-point correlations between qubits as a pairwise correlation matrix $\mathsf{Z}^{(p)}\in\mathbb{R}^{N\times N}$. Finally, we replace the adjacency matrix $\mathsf{W}$ with the correlation matrix $\mathsf{Z}^{(p)}$ in the problem's objective function. Hence, the quantum preconditioner preserves the structure of quadratic unconstrained binary optimization, and the correlation matrix is the quantum-preconditioned problem solved in place of the original problem using a suitable classical solver. This allows classical solvers to work with either the original input or a preconditioned version, potentially leading to a direct improvement over the current state-of-the-art classical approach.

Building on the adiabatic theorem, we demonstrate that in the adiabatic limit ($p\to +\infty$)~\cite{Farhi2000}, the quantum transformation renders solving the problem optimally a trivial task. Any local updates that improve a candidate solution $\boldsymbol{z}$ lead to the optimal solution $\boldsymbol{z}_{\textbf{opt}}$, thereby benefiting classical solvers like SA and parallel tempering. In the opposite limit, at a shallow circuit depth $p\leq 2$, we perform classical emulations of the QAOA to test the performance of the proposed quantum preconditioner. We find that best-in-class classical heuristics can converge more rapidly on an average case. In particular, SA and the BM algorithm can benefit from low-depth quantum preconditioning in solving a range of problems, such as Sherrington-Kirkpatrick spin glasses~\cite{PhysRevLett.35.1792}, random $3$-regular graph maximum-cut problems, and a real-world grid energy problem~\cite{ACTIVSg500}.

Using a light-cone decomposition technique~\cite{Dupont2024}, we benchmark the proposed approach on problems with thousands of variables and investigate the effect of increased circuit depth on its convergence. For instance, on an ensemble of random $3$-regular graph maximum-cut problems with $N=4,096$ variables, quantum preconditioning can make BM and SA converge one order of magnitude faster to a typical solution $\boldsymbol{z}$ with an average approximation ratio $\alpha=C(\boldsymbol{z})/C(\boldsymbol{z}_{\textbf{opt}})=99.9\%$ (Sec.~\ref{sec:practical_perf}). When accounting for the additional time taken for preconditioning, the benefit offered by shallow circuits remains and translates into a practical advantage for random $3$-regular graph maximum-cut problems through classical QAOA emulations. To the best of our knowledge, this is the first reported quantum-inspired advantage for optimization. The improved convergence observed upon increasing the circuit depth opens the door to a hardware-based quantum advantage using circuits that are beyond the reach of classical emulators for preconditioning. 

In addition, we study why quantum preconditioning makes the problem easier by considering small-$N$ problems at large $p$ (Appendix~\ref{app:ease_hardness}). We test an experimental implementation of quantum preconditioning of the real-world grid energy problem~\cite{ACTIVSg500} on a Rigetti Ankaa-3 superconducting device (Appendix~\ref{app:experiments_mpes}). Finally, we identify challenges and discuss prospects for quantum preconditioning to deliver utility (Sec.~\ref{sec:budget_prospects}). We give our conclusions in Sec.~\ref{sec:conclusion}.

\section{Quantum Preconditioning}
\label{sec:quantum_precond}

\begin{figure*}[!t]
    \centering
    \includegraphics[width=1\textwidth]{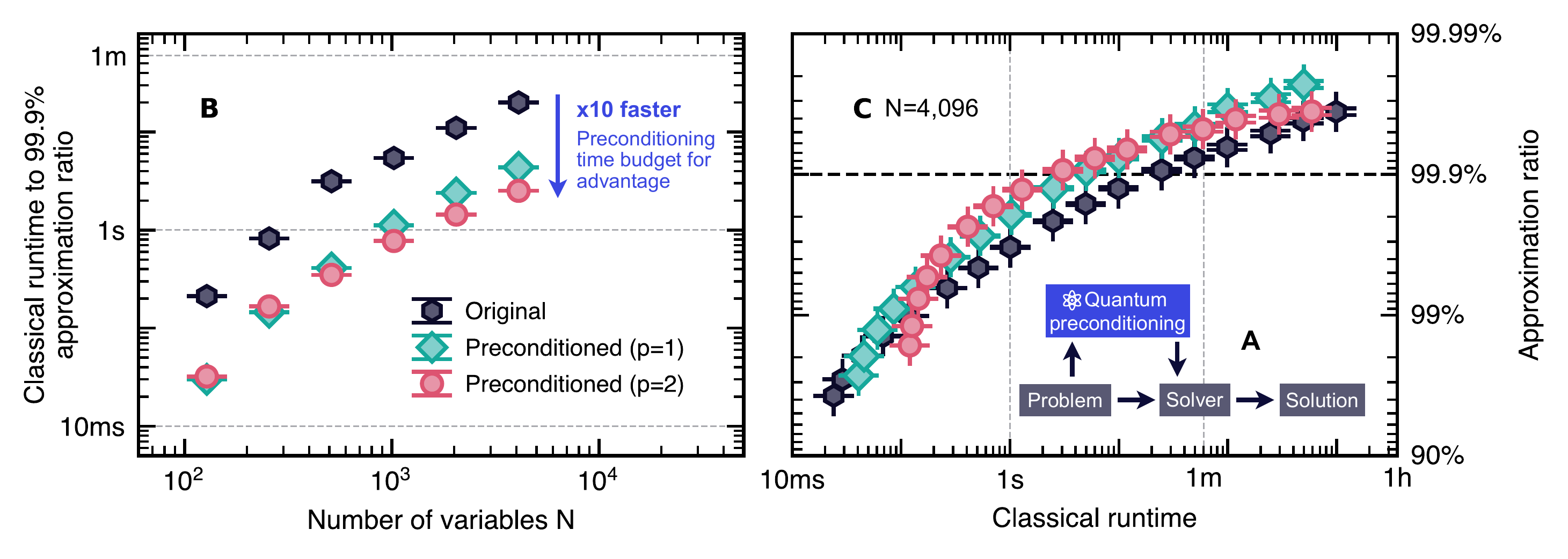}
    \caption{Average performance for the maximum-cut problem on $N$-variable random $3$-regular graphs via the classical simulated annealing (SA) solver based on the original and quantum-preconditioned problems. (A) Workflow diagram of quantum preconditioning. (B) Average run-time for SA to get to an approximation ratio of $99.9\%$ as a function of $N$. (C) Average approximation ratio versus SA run-time for $N=4,096$. Each data point is averaged over $200$ randomly generated problem instances. Run-times correspond to a 64GB MacBook Pro with an Apple M1 Max chip. Error bars indicate the standard error of the mean.}
    \label{fig:sa_preconditioning}
\end{figure*}

We consider discrete optimization problems of the form~\cite{lucas2014}
\begin{equation}
    C(\boldsymbol{z})=\frac{1}{2}\boldsymbol{z}^T\mathsf{W}\boldsymbol{z}=\frac{1}{2}\sum\nolimits_{i,j=1}^N\mathsf{W}_{ij}z_iz_j,
    \label{eq:obj_function}
\end{equation}
with variables $\boldsymbol{z}\in\{\pm 1\}^N$ and the symmetric matrix $\mathsf{W}\in\mathbb{R}^{N\times N}$ encoding the problem. The quantum preconditioner computes the two-point correlation matrix $\mathsf{Z}$ (see Fig.~\ref{fig:sa_preconditioning}a)
\begin{equation}
    \mathsf{W}_{ij}\leftarrow\mathsf{Z}_{ij}^{(p)} = \bigl(\delta_{ij} - 1\bigr)\bigl\langle\hat{Z}_i\hat{Z}_j\bigr\rangle_p,
    \label{eq:correlation_matrix}
\end{equation}
where $\delta_{ij}$ is the Kronecker delta and the expectation value involving Pauli-$Z$ operators is evaluated over the quantum state
\begin{equation}
    \bigl\vert\Psi\bigr\rangle_p=\left[\prod\nolimits_{\ell=1}^pe^{-i\beta_\ell\sum_{j=1}^N\hat{X}_j}e^{-i\gamma_\ell\hat{C}}\right]\hat{H}^{\otimes N}\vert{0}\rangle^{\otimes N},
    \label{eq:qaoa}
\end{equation}
where $\hat{H}$, $\hat{X}$, and $\langle\boldsymbol{z}\vert\hat{C}\vert\boldsymbol{z'}\rangle=C(\boldsymbol{z})\delta_{\boldsymbol{z}\boldsymbol{z'}}$ are the Hadamard gate, Pauli-$X$ operator, and operator encoding the objective function, respectively. There is a one-to-one mapping between a variable and a qubit. Equation~\eqref{eq:qaoa} corresponds to a standard $p$-layer QAOA circuit~\cite{Farhi2014,BLEKOS20241} parameterized via $\gamma_\ell$ and $\beta_\ell$, which are optimal when extremizing $\langle\hat{C}\rangle_p$. In a similar form to the original problem of Eq.~\eqref{eq:obj_function}, the preconditioned problem's objective function reads $\tilde{C}^{(p)}(\boldsymbol{z})=\frac{1}{2}\boldsymbol{z}^T\mathsf{Z}^{(p)}\boldsymbol{z}$. In the following, quantum preconditioning is performed at optimal or near-optimal $\gamma_\ell$ and $\beta_\ell$ values with respect to extremizing $\langle\hat{C}\rangle_p$ via the QAOA. Hence, the computational cost of preconditioning is the same as executing the QAOA. No additional information than what is available as part of the standard QAOA is needed.

\subsection{The asymptotic infinite-circuit-depth limit}

At infinite depth ($p\to+\infty$), the QAOA converges to the optimal solution~\cite{Farhi2014}. In the presence of a doubly degenerate optimal solution $\pm\boldsymbol{z}_\textbf{opt}$ due to the $\mathbb{Z}_2$ global sign-flip symmetry in the objective function of Eq.~\eqref{eq:obj_function}, the preconditioner takes the form $\mathsf{Z}^{(\infty)}=\mathsf{I}-\boldsymbol{z}_\textbf{opt}\boldsymbol{z}_\textbf{opt}^T$, where $\mathsf{I}$ is an $N\times N$ identity matrix. The objective function of the quantum-preconditioned problem now reads
\begin{equation}
    \tilde{C}^{(\infty)}(\boldsymbol{z}) = \frac{1}{2}\boldsymbol{z}^T\boldsymbol{z}-\vert\boldsymbol{z}_\textbf{opt}^T\boldsymbol{z}\vert^2,
    \label{eq:preconditioned_inf_obj_function}
\end{equation}
which is minimized for $\boldsymbol{z}=\pm\boldsymbol{z}_\textbf{opt}$. Therefore, both the original and preconditioned problems in the infinite-depth limit share the same optimal solution; however, while $\boldsymbol{z}_\textbf{opt}$ is nontrivial to find based on the original problem of Eq.~\eqref{eq:obj_function} in the general case, it can be straightforwardly obtained via local steps based on Eq.~\eqref{eq:preconditioned_inf_obj_function}. Indeed, $\tilde{C}^{(\infty)}(\boldsymbol{z}_\textbf{opt})=-N(N-1)$ marks the absence of frustration in the quantum-preconditioned problem. Each variable $z_i$ can be set to $\pm 1$ to locally minimize each term $\mathsf{Z}^{(\infty)}_{ij}z_iz_j$, which minimizes the objective function globally. This signals the absence of glassy properties~\cite{RevModPhys.58.801,Castellani_2005,charbonneau2023spin} in the quantum-preconditioned problem in the infinite-depth limit. Hence, we expect a strong benefit for solvers, such as SA~\cite{kirkpatrick1983optimization}, parallel tempering~\cite{hukushima1996exchange}, and others, that rely on local steps for convergence. This is evidenced by the decrease in frustration index of the preconditioned problem with QAOA circuit depth $p$, as shown in Appendix~\ref{app:ease_hardness} for small-$N$ problem instances.

To extend the infinite-depth properties of quantum preconditioning to problems beyond the doubly degenerate optimal solution $\pm\boldsymbol{z}_\textbf{opt}$, another preconditioner is needed. For instance, for problems with a unique solution $\boldsymbol{z}_\textbf{opt}$ due to an extra nonzero term of the form $\sum_ih_iz_i$ with $h_i\in\mathbb{R}$, one may use $\sum_ih_iz_i\leftarrow-\sum_i\langle\hat{Z}_i\rangle_p z_i$. For problems with degenerate optimal solutions beyond $\pm\boldsymbol{z}_\textbf{opt}$, one can add vanishing one-body terms $\lim_{h_i\to 0}h_iz_i$ in the original objective function to favor a single solution.

\subsection{At a shallow circuit depth}

The performance at practical circuit depths may be independent of these infinite-depth considerations. In the following, we investigate the performance of the preconditioner of Eq.~\eqref{eq:correlation_matrix} on problems with at least a doubly degenerate optimal solution $\pm\boldsymbol{z}_\textbf{opt}$ (potentially more) due to their global $\mathbb{Z}_2$ sign-flip symmetry. Here, the optimal solution for $\mathsf{Z}$ is generically different from the optimal solution for $\mathsf{W}$. Nevertheless, we find that this form of quantum preconditioning at shallow depth can still provide a benefit. Quantifying the quality of a solution $\boldsymbol{z}$ through the approximation ratio $\alpha=C(\boldsymbol{z})/C(\boldsymbol{z}_{\textbf{opt}})$, we find that the classical solvers achieve $\alpha\approx 1$ when working with the preconditioned problem, but they often do so with fewer iterations than with the original problem, which may translate to less computational time. This is the central result of this work. We emphasize that, although a solution $\boldsymbol{z}$ is obtained by solving the original or preconditioned problem, the objective function $C(\boldsymbol{z})$ and the optimal solution $\boldsymbol{z}_{\textbf{opt}}$ are always with respect to the original problem for evaluating the approximation ratio $\alpha$.

The intuition behind the QAOA preconditioner of Eq.~\eqref{eq:correlation_matrix} at a shallow circuit depth $p$ is that the correlation matrix contains similar information to the original problem. Indeed, locally, two variables $z_i$ will tend to take values such that $\mathsf{W}_{ij}z_iz_j$ is minimized, which translates into $\textrm{sign}(\mathsf{W}_{ij})\simeq -\textrm{sign}\langle\hat{Z}_i\hat{Z}_j\rangle$. Thus, one may anticipate that solving a problem based on $\mathsf{W}_{ij}$ or $\mathsf{Z}_{ij}^{(p)}$ will lead to similar solutions. A special case of quantum preconditioning is the quantum relax-and-round algorithm described in Refs.~\cite{PhysRevA.109.012429,Dupont2024}. There, it was a shown that a classical relax-and-round solver yields identical solutions for the original problem given by $\mathsf{W}$ and the preconditioned problem at $p=1$ given by $\mathsf{Z}^{(p=1)}$ on a range of problems, including Sherrington-Kirkpatrick spin glasses and random $3$-regular graph maximum-cut problems.

The same intuition holds true for any classical solver. Here, we employ the state-of-the-art SA and BM solvers to solve Sherrington-Kirkpatrick spin glasses~\cite{PhysRevLett.35.1792}, random $3$-regular graph maximum-cut problems, and a real-world grid energy problem~\cite{ACTIVSg500}. We find that at small $p$, these solvers converge faster to a given value of the approximation ratio $\alpha$ on the quantum-preconditioned versus the original problems, and conversely, they yield a higher approximation ratio $\alpha$ in a given amount of time. In addition, in Appendix~\ref{app:ease_hardness}, we investigate large-$p$ quantum preconditioning on small-$N$ problem instances.

\section{Practical Performance}
\label{sec:practical_perf}

\subsection{Random \texorpdfstring{$3$}{3}-regular graph maximum-cut problems}

\subsubsection{Classical simulated-annealing-solver baseline}

We investigate the average performance of SA~\cite{kirkpatrick1983optimization} on hundreds of random $N$-variable $3$-regular graph maximum-cut problems. SA is a physics-inspired solver treating an objective function such as Eq.~\eqref{eq:obj_function} as a classical Ising model from which one wants to find the ground state. The search is based on a Metropolis-Hastings Markov chain Monte Carlo algorithm with a Boltzmann distribution~\cite{10.1063/1.1699114,10.1093/biomet/57.1.97}. It starts at a high temperature, which is reduced throughout the iterations according to a predefined schedule such that one samples low-energy states, and ideally the ground state (see Appendix~\ref{app:classical_solvers}). The objective function of the maximum-cut problem is closely related to that of Eq.~\eqref{eq:obj_function}, as it seeks to maximize the cut number
\begin{equation}
    C_{\textrm{MaxCut}}(\boldsymbol{z})=\frac{1}{4}\sum\nolimits_{i,j=1}^N\mathsf{W}_{ij}\Bigl(1-z_iz_j\Bigr),
    \label{eq:obj_function_maxcut}
\end{equation}
where $W_{ij}=1$ if there is an edge between vertices $i$ and $j$ and zero otherwise. In Fig.~\ref{fig:sa_preconditioning}c, we evaluate the average approximation ratio $\alpha$ as a function of the solver's run-time for $N=4,096$ variables. We estimate and report the run-time to get to $\alpha=99.9\%$ as a function of $N$ in Fig.~\ref{fig:sa_preconditioning}b.

\subsubsection{Performing quantum preconditioning}
\label{sec:light_cone_3reg}

Having defined a classical baseline, we now quantum precondition the problems using exact state-vector emulations. We employ a light-cone decomposition technique~\cite{Dupont2024} to access two-point correlations $\langle\hat{Z}_i\hat{Z}_j\rangle_p$ of Eq.~\eqref{eq:correlation_matrix} on problems much larger than qubits na\"ively accessible.

A $p$-layer QAOA circuit induces quantum coherence around each qubit and its $p$-nearest neighbors. Therefore, one can estimate a two-point correlation between two qubits $i$ and $j$ via the QAOA by only considering the subset of qubits resulting from the intersection of their respective light cones. If the light cones do not intersect, then $\langle\hat{Z}_i\hat{Z}_j\rangle_p=\langle\hat{Z}_i\rangle_p\langle\hat{Z}_j\rangle_p=0$ due to the global sign-flip symmetry of Eq.~\eqref{eq:obj_function_maxcut}. The topology of $3$-regular graphs is such that intersecting light cones have at most $1+6(2^p-1)$ qubits, making $p\leq 2$ readily accessible through standard circuit-emulation methods.

Moreover, because the average distance between two vertices grows as $O(\ln N)$ in $3$-regular graphs, only $O(N)$ light cones will intersect in the limit $p\ll\ln N$, out of the $O(N^2)$ na\"ive possibilities. In summary, we trade a single $N$-qubit QAOA circuit for $O(N)$ QAOA circuits with $O(\exp p)$ qubits each. Variational parameters are set to near-optimal tabulated values~\cite{PhysRevA.103.042612,Wurtz2021}.

\begin{figure}[!t]
    \centering
    \includegraphics[width=1\columnwidth]{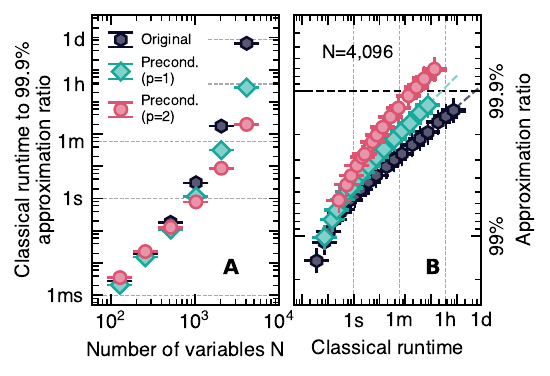}
    \caption{Average performance for the maximum-cut problem on $N$-variable random $3$-regular graphs via the classical BM solver based on the original and quantum-preconditioned problems. (A) Average run-time for BM to get to an approximation ratio of $99.9\%$ as a function of $N$. (B) Average approximation ratio versus BM run-time for $N=4,096$. Each data point is averaged over $200$ randomly generated problem instances. Run-times correspond to a 64GB MacBook Pro with an Apple M1 Max chip. Error bars indicate the standard error of the mean.}
    \label{fig:burer2002_preconditioning}
\end{figure}

Quantum preconditioning increases the number of nonzero terms $n$ in the problem because correlations extend beyond nearest-neighbor: There are $n=1.5N$ terms in the original $3$-regular graph problem, $n\simeq 4.5N$ at $p=1$, and $n\simeq 22.5N$ at $p=2$ in the large-$N$ limit (see Appendix~\ref{app:n_nonzero_terms_3reg}). Although classical solvers typically have a run-time proportional to the number of terms $n$ (see Appendix~\ref{app:iterations_to_run-time}), we show in the following that the increase in the number of terms is not a bottleneck for quantum preconditioning to deliver an advantage.

\subsubsection{Quantum-preconditioned simulated annealing solver}

In Fig.~\ref{fig:sa_preconditioning}, we report the performance of quantum preconditioning on SA for hundreds of random $3$-regular graph maximum-cut problems. We find that SA gets to an approximation ratio of $99.9\%$ about one order of magnitude faster when working with the quantum-preconditioned input rather than the original problem. For the largest problem sizes considered, we also observe a separation between $p=1$ and $p=2$, where increasing the circuit depth further accelerates the convergence of the classical solver. At $p\to+\infty$, we anticipate that SA will trivially converge to the optimal solution.

At a fixed number of variables ($N=4,096$), Fig.~\ref{fig:sa_preconditioning}c shows that quantum preconditioning can deliver a benefit across a range of approximation-ratio and run-time targets. As the run-time increases, we observe that SA on the $p=2$ quantum preconditioned-problem saturates to an approximation ratio $\alpha<1$. This is because, as mentioned earlier, at finite $p$, the optimal solutions of the original and quantum-preconditioned problems are generically different; however, it is interesting to see that in practice, their respective optimal solutions are spectrally close with the solver attaining $\alpha\simeq 99.96\%$ at $p=2$. We provide additional data for other values of $N$ in Appendix~\ref{app:perf_niter}.

\subsubsection{Quantum-preconditioned Burer-Monteiro solver}

Next, we consider quantum preconditioning in the context of the BM solver~\cite{burer2002rank,Burer2003}. The BM solver (see Appendix~\ref{app:classical_solvers}) is inspired by semidefinite programming methods: it relaxes the discrete nature of the variables in the objective function, solves this modified problem, and rounds the solution back to the valid domain at the end. In practice, the BM algorithm is one of the best heuristics to date for the maximum-cut problem~\cite{DunningEtAl2018}. Depending on the problem size and the performance target, it may be a better choice than SA.

We observe in Fig.~\ref{fig:burer2002_preconditioning} that quantum preconditioning provides a run-time advantage to get to an average approximation ratio of $\alpha=99.9\%$ on the largest problem sizes. For example, for $N=4,096$, the BM solver takes only a few minutes to solve the preconditioned problem at $p=2$ and yield $\alpha=99.9\%$, a notable improvement over one hour required at $p=1$, and about one day to solve the original problem. Although SA is more efficient than the BM solver in the $\alpha=99.9\%$ regime, including via quantum preconditioning, the hierarchy may change for a larger circuit depth $p$.

The performance boost offered by quantum preconditioning and observed on solvers as different as SA and BM suggests the wide applicability of the technique. In particular, the quantum relax-and-round solver~\cite{PhysRevA.109.012429,Dupont2024} is another example of quantum preconditioning via a classical relax-and-round solver inspired by semidefinite programming methods. In that case, one can prove that the performance at $p=1$ matches that of the classical counterpart and increases with $p$ on classes of problems---including Sherrington-Kirkpatrick spin glasses and random $3$-regular graph maximum-cut problems~\cite{PhysRevA.109.012429,Dupont2024}.

\subsection{Sherrington-Kirkpatrick spin glasses}

\subsubsection{Problem and quantum preconditioning}

We now turn our attention to another class of problems: Sherrington-Kirkpatrick (SK) spin glasses~\cite{PhysRevLett.35.1792}. The goal is to minimize an objective function of the form of Eq.~\eqref{eq:obj_function}, where the adjacency matrix $\mathsf{W}$ encoding the problem is drawn from the Gaussian orthogonal ensemble. Namely, its entries are independent and identically distributed random variables from a normal distribution of zero mean and unit width $\mathsf{W}_{ij}\sim\mathcal{N}(0,1)$, with $\mathsf{W}_{ij}=\mathsf{W}_{ji}$. SK problem instances are quite different from the random $3$-regular graph maximum-cut problems considered previously and constitute another standard optimization benchmark~\cite{dupont2023quantum,Farhi2022,PhysRevApplied.22.044074,Harrigan2021,venturelli2015quantum}. First, they are dense, with $O(N^2)$ terms instead of $O(N)$ for $N$-variable problems. Second, they are weighted problems with $\mathsf{W}_{ij}\neq 1$.

We perform quantum preconditioning at $p=1$, where one can compute semianalytically the correlation matrix of Eq.~\eqref{eq:correlation_matrix} via a back-propagation technique~\cite{PhysRevA.109.012429} (reproduced in Appendix~\ref{app:corr_formula} for completeness). We use optimal QAOA angles $\gamma=1/2\sqrt{N}$ and $\beta=\pi/8$ from the $N\to+\infty$ limit~\cite{Farhi2022}, thus skipping variational optimization. Unlike the $3$-regular graph maximum-cut problems, the resulting quantum-preconditioned SK spin glasses conserve the number of terms and the same statistics as the original $\mathsf{W}_{ij}$~\cite{PhysRevA.109.012429}. Emulating the QAOA at $p>1$ for large $N$ is currently not feasible.

\subsubsection{Performance}

\begin{figure}[!t]
    \centering
    \includegraphics[width=1\columnwidth]{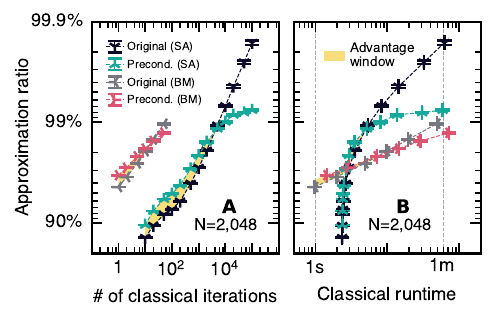}
    \caption{Average performance for $N=2,048$ Sherrington-Kirkpatrick spin-glass problems via the SA and Burer-Monteiro (BM) solvers based on the original and quantum-preconditioned $(p=1)$ problems. (A) Average approximation ratio as a function of the number of iterations. (B) Average approximation ratio as a function of the run-time. Each data point is averaged over $200$ randomly generated problem instances. Run-times correspond to a 64GB MacBook Pro with an Apple M1 Max chip. Error bars indicate the standard error of the mean.}
    \label{fig:sk_preconditioning}
\end{figure}

In Fig.~\ref{fig:sk_preconditioning}, we show the performance of SA and the BM solver on $200$ randomly generated SK instances with $N=2,048$ variables (additional data at other values of $N$ are available in Appendix~\ref{app:perf_niter}). In particular, we plot the approximation ratio as a function of the number of iterations (Fig.~\ref{fig:sk_preconditioning}a) and run-time (Fig.~\ref{fig:sk_preconditioning}b) with the relation between the two discussed in Appendix~\ref{app:iterations_to_run-time}.

\begin{figure*}[!t]
    \centering
    \includegraphics[width=1\textwidth]{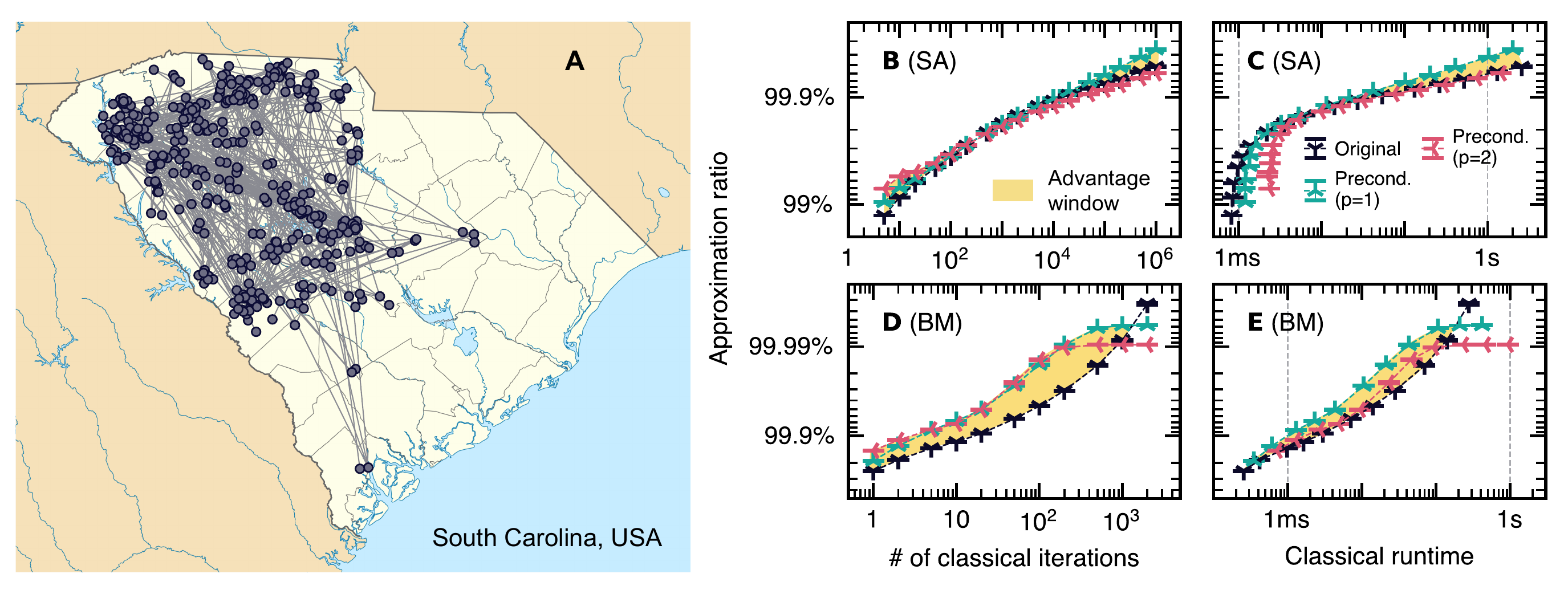}
    \caption{Performance of quantum preconditioning for computing the maximum power exchange section of an energy-grid optimization problem. (A) Representation of the energy-grid optimization problem considered from the state of South Carolina in the United States of America. Vertices ($N=180$) and edges ($n=226$) correspond to buses and lines, respectively. (B) Average approximation ratio as a function of the number of iterations via SA. (C) Average approximation ratio as a function of run-time via SA. (D) Average approximation ratio as a function of the number of iterations via the BM solver. (E) Average approximation ratio as a function of run-time via the BM solver. Run-times are based on a 64GB MacBook Pro with an Apple M1 Max chip. Each data point is averaged over $10^4$ samples. Error bars indicate the standard error of the mean.}
    \label{fig:mpes_preconditioning}
\end{figure*}

We observe that for a large number of iterations (or run-time), the approximation ratio obtained via the quantum-preconditioned problem saturates to $\alpha\simeq 99.3\%$. This, again, indicates that it does not share the same optimal solution as the original problem. For a small number of iterations, we find that the classical solver benefits from working with the preconditioned problem, achieving a higher approximation ratio at the same iteration count. This leads to a crossing of performance and an advantage window for quantum preconditioning on SK problems for both SA and BM solvers.

The overhead of about two seconds for the SA data in Fig.~\ref{fig:sk_preconditioning}b originates from the $O(N^2)$ heuristic used for setting the temperature schedule (see Appendix~\ref{app:classical_solvers}). It is absent in the case of the BM solver, which does not have any tuned hyperparameters. The crossing of performance happens at slightly different approximation ratio values for SA and the BM algorithm. This can be understood as both solvers returning different distributions of solutions.

\subsection{Power-grid optimization problem: The maximum power exchange section}
\label{sec:grid_opt_problem}

\subsubsection{Problem definition}

The last example considered is a real-world power-grid optimization problem seeking to compute the maximum power exchange section (MPES) of an energy network~\cite{Jing2023,9813820,9948058,Bauer2024}. The MPES is a metric that provides information about the health and the power-delivery capability of the energy grid. We consider the realistic $500$-bus power system dataset \textit{ACTIVSg500}, which mimics the energy grid of the state of South Carolina in the United States of America~\cite{ACTIVSg500}; see Fig.~\ref{fig:mpes_preconditioning}a. The dataset was designed as part of the \textit{ARPA-E's GRID DATA} program to be similar to the actual electric grid while containing no confidential information about critical energy infrastructure~\cite{7459256,7515182,7725528,xu2017creation,xu2017}. 

The MPES is a weighted maximum-cut problem. As such, it seeks to maximize the objective function
\begin{equation}
    C_{\textrm{MPES}}(\boldsymbol{z})=\frac{1}{4}\sum\nolimits_{i,j=1}^N\mathsf{W}_{ij}\Bigl(1-z_iz_j\Bigr),
    \label{eq:obj_function_mpes}
\end{equation}
where the weight $\mathsf{W}_{ij}$ is given by the line impedance between two buses $i$ and $j$
\begin{equation}
    \mathsf{W}_{ij}=\left(\mathsf{R}_{ij}^2+\mathsf{X}_{ij}^2\right)^{-1/2},
    \label{eq:line_impedance}
\end{equation}
with $\mathsf{W}_{ji}=\mathsf{W}_{ij}$ and $\mathsf{W}_{ij}=0$ if buses $i$ and $j$ are not connected. Further, $\mathsf{R}_{ij}$ and $\mathsf{X}_{ij}$ correspond to the line resistance and reactance, respectively. Their values are provided as part of the dataset.

The original problem contains $N=410$ variables and $n=463$ terms. The corresponding graph problem contains dangling tree branches, which we remove. Because their topology is nonfrustrating, it is trivial to find the optimal value for the variables on those branches once a solution for the rest of the problem is found: one simply sets the value of a variable $z_j=\pm 1$ such that $\mathsf{W}_{ij}z_iz_j$ is locally minimized. The remaining problem has $N=201$ variables and $n=254$ terms; however, one finds that it actually consists of two disconnected and independent subgraphs. The first contains $N=180$ nodes and $n=226$ edges; the second contains $N=21$ nodes and $n=28$ edges and can, therefore, be solved via brute-force enumeration. In the following, we solve the larger of the two subproblems and trivially reconstruct a solution $\boldsymbol{z}$ to the original problem of Eq.~\eqref{eq:obj_function_mpes}.

The optimal contribution of the dangling tree branches to the global objective value is approximately $50,778.093880$, the optimal contribution of the $N=21$ nodes problem to the global objective value is approximately $3,728.413221$, and we estimate that the global optimal solution has an objective value $C_{\textrm{MPES}}(\boldsymbol{z}_\textbf{opt})\simeq 61,606.078827$.

\subsubsection{Quantum preconditioning}

The $N=180$ graph problem is sufficiently sparse that we can employ the light-cone technique introduced in the context of the random $3$-regular graph maximum-cut problems in Sec.~\ref{sec:light_cone_3reg}. The graph has an average and maximum degree of approximately $2.51$ and $7$, respectively. At $p=1$ and $p=2$, the light-cone-induced subgraphs for building the correlation matrix have at most $14$ and $27$ vertices, respectively. Both numbers are within the ability of classical state-vector emulators, which we use for preconditioning.

Unlike the previous graph problems investigated, near-optimal QAOA angles are not tabulated for this problem. As such, we perform a variational search for the QAOA angles extremizing the expectation value of the objective function $\langle\hat{C}\rangle_p$. We use the Broyden-Fletcher-Goldfarb-Shanno (BFGS)~\cite{BFGS1,BFGS2,BFGS3,BFGS4} algorithm, starting from random initial angles a few hundred times and consider the best output for computing the correlation matrix. We report the QAOA angles in Appendix~\ref{app:problem_instances}. The original and preconditioned problems have $n=226$, $n=600$ $(p=1)$, and $n=1,625$ $(p=2)$ terms, respectively.

\subsubsection{Performance}

We display the average approximation as a function of the number of iterations and run-time for SA and the BM solver in Fig.~\ref{fig:mpes_preconditioning}. We compare the performance of these solvers on the original and preconditioned problems. We evaluate the average approximation ratio by sampling $10^4$ solutions from the classical solvers for each number of iterations. We observe that for this problem in the range considered, the BM algorithm is more efficient than SA as it can reach a much higher approximation ratio for the same fixed run-time.

For a given number of iterations, we find regions where the solvers on the preconditioned problems yield a higher approximation ratio than with the original one. This defines windows for an advantage. At a small number of iterations, we observe that the window is enlarged from $p=1$ to $p=2$. Such a gain from increasing the amount of preconditioning was also observed for the random $3$-regular graph maximum-cut problems. When substituting the number of iterations for the run-time on a 64GB MacBook Pro with an Apple M1 Max chip, there is still an advantage window, despite the quantum-preconditioned problems having more terms than the original one (see Appendix~\ref{app:n_nonzero_terms_3reg}). Finally, we note that $p=1,2$ preconditioned problems can access approximation ratios $\alpha\gtrsim 99.99\%$ and that they do not share the same optimal solutions as the original problem as exemplified by the saturation in approximation ratio observed in Fig.~\ref{fig:mpes_preconditioning}d. In addition to classical state-vector emulations, we precondition the problem using an experimental implementation of the QAOA at $p=1$ and solve the preconditioned problem. Results of this experiment are shown in App.~\ref{app:experiments_mpes}.

The existence of an advantage window for a real-world problem highlights the potential wide applicability of the proposed quantum-preconditioning method.

\section{Quantum-Preconditioning Budget and Prospects for a Quantum Advantage}
\label{sec:budget_prospects}

We have shown that quantum-preconditioning problems can benefit state-of-the-art classical solvers. While this is a nontrivial result on its own, a practical advantage can only be claimed if one also accounts for the preconditioning time itself---see, e.g., Fig.~\ref{fig:sa_preconditioning}b defining a preconditioning time budget for random $3$-regular graph maximum-cut problems via SA.

\subsection{Quantum-inspired advantage for random \texorpdfstring{$3$}{3}-regular graph maximum-cut problems}

\begin{figure}[!t]
    \centering
    \includegraphics[width=1\columnwidth]{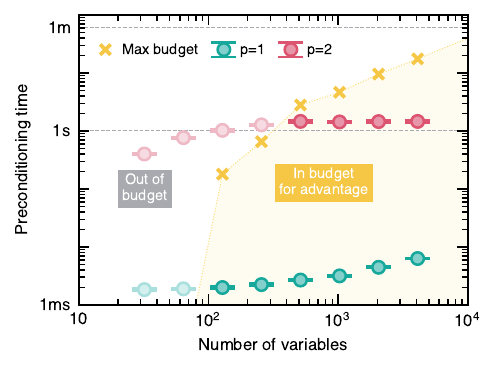}
    \caption{Average time for preconditioning $N$-variable random $3$-regular graph maximum-cut problems using classical state-vector emulations of the QAOA using the light-cone technique at $p=1$ and $p=2$. The maximum preconditioning time budget for an advantage via SA for an approximation ratio of $\alpha=99.9\%$ is computed from Fig.~\ref{fig:sa_preconditioning}. Run-times correspond to a 64GB MacBook Pro with an Apple M1 Max chip. Error bars indicate the standard error of the mean.}
    \label{fig:runtime_preconditioning}
\end{figure}

We refer to a quantum-inspired advantage as obtained via a purely classical calculation or emulation of a quantum circuit, as opposed to execution on an actual quantum computer. This can be a first step toward genuine quantum utility if the proposed method has room to perform even better at a larger scale (more qubits, larger depth) beyond classical computational capabilities.

We focus on the random $3$-regular graph maximum-cut problems, which present the largest advantage window compared to the SK spin glasses or the grid energy problem. We consider SA, which is more efficient than the BM solver in the $\alpha=99.9\%$ regime (this value has been chosen arbitrarily, and others may provide a larger advantage window). The preconditioning time budget is defined in Fig.~\ref{fig:sa_preconditioning}b as the difference in SA run-time between the original and preconditioned problems to reach $\alpha=99.9\%$ . We report the budget between original and preconditioned at $p=2$ in Fig.~\ref{fig:runtime_preconditioning}. Because there is a difference of about one order of magnitude in runtime between the original and preconditioned problems, the budget is mostly dominated by the run-time of the solver on the original problem.

We perform the preconditioning via state-vector emulations of the QAOA circuits on the relevant light-cone-induced subgraphs according to Sec.~\ref{sec:light_cone_3reg}. Determining the light-cone-induced subgraphs of an $N$-variable problem scales as $O(N)$. The QAOA implementation benefits from the global $\mathbb{Z}_2$ sign-flip symmetry of the problem, which reduces the overall emulation run-time by a theoretical factor two. Moreover, we use the fact that in the limit $p\ll\ln N$, most light-cone-induced subgraphs are trees (see Appendix~\ref{app:n_nonzero_terms_3reg}). Trees are $N$-variable graphs for which the number of edges is $n=N-1$. As such, we cache the expectation value $\langle\hat{Z}_i\hat{Z}_j\rangle$ of the trees such that the QAOA only needs to be emulated once on such subgraphs. This results in only emulating $O(\exp p)$ QAOA circuits, independent of $N$ (see Appendix~\ref{app:n_nonzero_terms_3reg}). The memory usage and algorithmic complexity of a QAOA emulation via the light-cone technique scale as $O(\exp\exp p)$ and are classically manageable at $p=1$ and $p=2$.

\begin{figure*}[!t]
    \centering
    \includegraphics[width=1\textwidth]{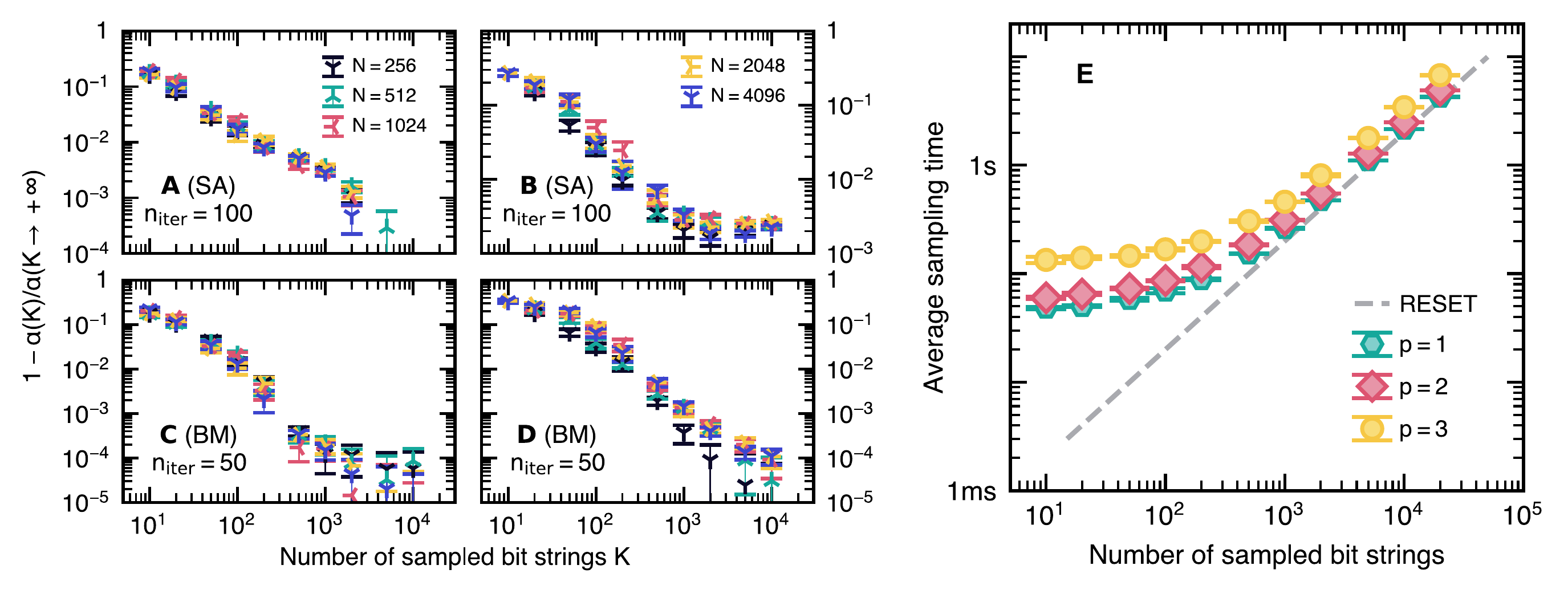}
    \caption{(A-D) Random $N$-variable $3$-regular graph maximum-cut problems are considered, showing the convergence of the average approximation ratio $\alpha$ toward its infinite-sampling value, obtained via quantum preconditioning with a finite number of sampled bit strings $K$. Preconditioning is performed via the simulation of light-cone-induced $p$-layer QAOA circuits. (A) SA at $p=1$ with $n_\textrm{iter}=100$ iterations. (B) SA at $p=2$ with $n_\textrm{iter}=100$ iterations. (C) BM solver at $p=1$ with $n_\textrm{iter}=50$ iterations. (D) BM solver at $p=2$ with $n_\textrm{iter}=50$ iterations. Each data point is averaged over $200$ randomly generated problem instances. (E) Average sampling time on Rigetti Ankaa-3 superconducting quantum chip for a light-cone-induced QAOA circuit for various numbers of QAOA layers $p$ for random $3$-regular graph maximum-cut problems. $\texttt{RESET}$ corresponds to a passive reset time of $200~\mu$s of the qubits between two circuit executions. Each data point is averaged over $50$ randomly selected light-cone-induced QAOA circuits. Error bars indicate the standard error of the mean.}
    \label{fig:preconditioning_nshots}
\end{figure*}

In Fig.~\ref{fig:runtime_preconditioning}, we show the time it takes to perform the quantum preconditioning at $p=1$ and $p=2$ via emulations. For large enough problems, the preconditioning run-time fits within the budget. Therefore, quantum preconditioning can deliver a quantum-inspired advantage over SA through classical emulations. This advantage extends for larger problem sizes to the BM solver, which has a much larger budget (see Fig.~\ref{fig:burer2002_preconditioning}); however, the BM solver is not competitive against SA.

The change of slope observed at large $N$ for $p=1$ is because the run-time becomes dominated by the $O(N)$ task of determining the light-cone-induced subgraphs rather than the QAOA emulations. We anticipate an analogous asymptotic behavior for $p=2$.

\subsection{Prospects for hardware-based quantum utility for random \texorpdfstring{$3$}{3}-regular graph maximum-cut problems}
\label{sec:propsects_3reg}

We have shown in Fig.~\ref{fig:runtime_preconditioning} that quantum preconditioning can provide a quantum-inspired advantage on random $3$-regular graph maximum-cut problems via classical emulations at $p=1$ and $p=2$. We have also shown in Fig.~\ref{fig:sa_preconditioning}b that increasing the depth of the QAOA from $p=1$ to $p=2$ can accelerate the convergence. Hence, a question is whether $p=3$ could improve the convergence further; however, light-cone-induced subgraphs at $p=3$ have up to $43$ qubits and, therefore, cannot be emulated with exact classical methods in a reasonable amount of time, calling for a new simulation platform, such as an actual quantum computer.

\subsubsection{Sampling from the wave function}

When moving to quantum hardware for executing quantum circuits, one does not have access to the full quantum state $\vert\Psi\rangle_p$ of Eq.~\eqref{eq:qaoa} for computing expectation values but must sample bit strings $\{\boldsymbol{z}^{(k)}\}_{k=1,\ldots K}$ to estimate them. After executing the circuit and measuring the qubits $K$ times, one has collected $K$ bit strings according to the probability $\vert\langle\boldsymbol{z}^{(k)}\vert\Psi\rangle_p\vert^2$. Two-point correlations are then estimated as
\begin{equation}
    \bigl\langle\hat{Z}_i\hat{Z}_j\bigr\rangle_p\simeq\frac{1}{K}\sum\nolimits_{k=1}^Kz_i^{(k)}z_j^{(k)}.
    \label{eq:correlation_via_bitstrings}
\end{equation}
The total run-time for estimating the expectation value is proportional to $K$. For the classical state-vector emulations, we had access to the full quantum state $\vert\Psi\rangle_p$ and did not need to sample bit strings, which corresponded to the $K\to+\infty$ limit.

\paragraph{Performance against sampling---}

A first question is whether quantum preconditioning is robust to finite sampling. In Figs~\ref{fig:preconditioning_nshots}A-~\ref{fig:preconditioning_nshots}D, we show the accuracy of SA and the BM solver on preconditioned $N$-variable random $3$-regular graph maximum-cut problems where preconditioning was performed by sampling $K$ bit strings to estimate correlations via Eq.~\eqref{eq:correlation_via_bitstrings}. The correlation matrix was computed via the classical emulation of light-cone-induced QAOA subgraphs, as was done previously in Sec.~\ref{sec:light_cone_3reg}. We consider preconditioning at $p=1$ and $p=2$ and the performance of SA and BM with $n_\textrm{iter}=100$ and $n_\textrm{iter}=50$ iterations, respectively.

We do not observe a clear $N$ or $p$ dependence of the convergence of the average approximation ratio toward its infinite-sampling ($K\to+\infty$) value; however, we can rule out a logarithmic convergence with $K$ (i.e., an exponential number of samples needed to achieve a desired accuracy). Moreover, we note that a convergence to a relative error in average approximation ratio of $0.01\%$ requires at most $K\simeq 10^4$ bit strings, which is a typical number used for evaluating expectation values on quantum computers. The saturation observed at $K\sim 10^4$ in Figs.~\ref{fig:preconditioning_nshots}B and~\ref{fig:preconditioning_nshots}C is due to finite statistics in the number of random problem instances considered and for estimating the approximation ratio of the individual problem instances. We expect all the curves to get asymptotically to zero as $K\to +\infty$. Therefore, finite sampling does not seem to be a barrier to the performance of quantum preconditioning.

\paragraph{Sampling time---}
\label{sec:sampling_time}

The next question is the time needed for preconditioning via sampling. We consider the superconducting quantum computer Ankaa-3 developed by Rigetti Computing and accessed via cloud services. Superconducting qubits provide the fastest sampling rate to date thanks to their logical quantum operations in the tens-of-nanoseconds range, several orders of magnitude faster than atom- or ion-based quantum computers. Given that optimization solvers operate on an accuracy versus run-time window, the speed of preconditioning is paramount to deliver any potential advantage.

In Fig~\ref{fig:preconditioning_nshots}E, we report the average sampling time for the $p$-layer QAOA on a typical $N$-vertex subgraph induced by the light cone on $3$-regular graphs. An arbitrary QAOA circuit on $N$ qubits with $p$ layers executed on a linear chain topology of qubits through a na\"ive swap network~\cite{OGorman2019} requires a total of $3Np$ compact layers of two-qubit $\texttt{ISWAP}=\exp[i\pi(\hat{X}\otimes\hat{X} + \hat{Y}\otimes\hat{Y})/4]$ gates and $2(2N+1)p+1$ compact layers of one-qubit $\texttt{RX}(\phi)=\exp(-i\hat{X}\phi/2)$ gates where $\phi=\pm\pi/2$~\cite{Dupont2024}. $\hat{X}$, $\hat{Y}$, and $\hat{Z}$ are Pauli operators. All the gates within a layer can be applied in parallel on independent qubits. We also make use of one-qubit $\texttt{RZ}(\theta\in\mathbb{R})=\exp(-i\hat{Z}\theta/2)$ gates, which are virtual and realized by a change of frame of the qubits tracked by the control system. The collection time of $K$ samples is modeled by
\begin{equation}
    t\bigl(K, N, p\bigr) = t_\textrm{ovhead}\bigl(K, N, p\bigr) + Kt_\textrm{circ}\bigr(N, p\bigr),
    \label{eq:sampling_time}
\end{equation}
where $t_\textrm{ovhead}(K, N, p)$ indicates a classical overhead encompassing, e.g., network latency, job management, and control systems. The other component,
\begin{equation}
    t_\textrm{circ}\bigl(N, p\bigr) = 3Npt_\textrm{2Q} + \bigl[2(2N+1)p+1\bigr]t_\textrm{1Q} + t_\textrm{mes} + t_\textrm{res},
    \label{eq:circuit_time}
\end{equation}
relates to the circuit execution. It defines a minimum physical limit for sampling one quantum circuit. We have $t_\textrm{2Q}\simeq 80$ ns as the typical duration of a two-qubit gate, $t_\textrm{1Q}\simeq 40$ ns as the typical duration of a one-qubit gate, $t_\textrm{mes}=1~\mu$s as the typical duration of measuring a qubit, and $t_\textrm{res}=200~\mu$s as the typical duration for passively reseting a qubit prior to a subsequent circuit execution. These numbers provide a baseline, but they can typically be calibrated or fine-tuned.

For $N=4,096$, the average size of light-cone-induced subgraphs is $N=5.67(2)$ at $p=1$, $N=16.613(5)$ at $p=2$, and  $N=39.772(5)$ at $p=3$. Hence, in this regime, the average circuit execution time $t_\textrm{circ}$ is dominated by the passive reset time of the qubits. This is highlighted in Fig.~\ref{fig:preconditioning_nshots}E, where for a large number of samples, $t\simeq Kt_\textrm{res}$. For a small number of samples, the sampling time is instead dominated by the classical overhead. For a number of samples $K\gtrsim 10^3$, a first step for improving the sampling time is, therefore, to move to an active reset strategy for the qubits, which would yield $t_\textrm{mes} + t_\textrm{res}\simeq 6~\mu$s~\cite{Karalekas_2020}, a $30$-fold reduction over passive reset.

At the moment, for $K=10^4$, the total sampling time for a single light-cone-induced circuit takes about two to three seconds, as shown in Fig~\ref{fig:preconditioning_nshots}E. This is not competitive against classical state-vector emulations at $p=1$ and $p=2$, which can precondition \textit{all} light-cone-induced circuits in less time (see Fig.~\ref{fig:runtime_preconditioning}). Thus, improvements will be needed to reduce the sampling time and make the approach competitive and capable of delivering an advantage when accounting for the preconditioning time budget. Whether this will prove a bottleneck remains to be established: indeed, we observe in Fig.~\ref{fig:runtime_preconditioning} that the budget increases with the number of variables $N$ while the preconditioning is independent of $N$ for $p\ll \ln N$ thanks to caching of light-cone-induced subcircuits (see Appendix.~\ref{app:n_nonzero_terms_3reg})---strictly speaking, preconditioning has a classical $O(N)$ dependence in finding the light-cone-induced subgraphs. If confirmed for much larger $N$, this scaling difference between the preconditioning budget and time could suggest a crossing value $N^*$ for a quantum simulation method (hardware or classically emulated) to deliver an advantage, independently of its intrinsic run-time.

\subsubsection{Performance of a noisy quantum computer}
\label{sec:noisy_perf}

In the absence of quantum error correction, inherent hardware noise is another consideration for contemporary devices. To get a sense of the effect of noise through a back-of-the-envelope calculation, we assume that two-qubit gates are the dominant source of errors with an average fidelity $f$. An arbitrary QAOA circuit on $N$ qubits with $p$ layers executed on a linear chain topology of qubits through a na\"ive swap network requires $n_\textrm{2Q}=3pN^2/2$ two-qubit gates~\cite{OGorman2019}. We have assumed three hardware-native two-qubit gates (such as $\texttt{ISWAP}$ or $\texttt{CZ}$) for compiling each two-qubit gate of the QAOA circuit. We suppose a fully depolarizing noise model parameterized by $F\in[0, 1]$, which results in the mixed state $\hat{\rho}_{p,F}=F\vert\Psi\rangle\langle\Psi\vert_p+(1-F)\hat{I}/2^N$, where $\hat{I}$ is the identity matrix. Two-point correlations of interest read
\begin{equation}
    \bigl\langle\hat{Z}_i\hat{Z}_j\bigr\rangle_{p,F}=\textrm{tr}\bigl(\hat{Z}_i\hat{Z}_j\hat{\rho}_{p,F}\bigr)=F\bigl\langle\hat{Z}_i\hat{Z}_j\bigr\rangle_{p,F=1},
    \label{eq:noise_rescaling}
\end{equation}
where $F=f^{n_\textrm{2Q}}$ can be interpreted as the global quantum circuit fidelity, with individual errors happening independently and at random at the two-qubit gate level. To distinguish an $O(1)$ expectation value with $K$ sampled bit strings, one needs $K\sim F^{-2}$, or equivalently $K\sim f^{-3pN^2}$, according to the central limit theorem. By plugging in relevant numbers for $f$, $N$, and $p$, one realizes that finite sampling by itself is likely not a bottleneck, but sampling for reliably estimating noisy and exponentially small expectation values might be. Therefore, hardware performance through $f$ sets the expectation for what circuit can be successfully executed. We verify that such a depolarizing noise model is able to phenomenologically fit experimental data in App.~\ref{app:experiments_mpes}, as similarly observed in Ref.~\cite{Harrigan2021}.

\subsubsection{Noisy quantum computer versus approximate classical tensor networks}
\label{sec:approx_tn_vs_nisq}

Given the fidelity requirements to run the QAOA on large problem instances in the absence of quantum error correction, one will likely need to rely on a decomposition technique such as the light-cone one introduced in Sec.~\ref{sec:light_cone_3reg}. Light-cone-induced subgraphs at $p=3$ require at most $43$ qubits, far beyond what state-vector emulators can handle in a reasonable amount of time. Other candidate simulators for simulating such a quantum circuit include an actual quantum computer or classical approximate tensor-network methods, such as matrix product states~\cite{ORUS2014117,RevModPhys.93.045003}. In the absence of quantum error correction, quantum computers are imperfect, and thus provide only an approximate simulation in some respects. A first question is whether quantum preconditioning via an approximately classically emulated preconditioner can deliver an advantage (see Appendix~\ref{app:perf_approx_mps}).

Assuming an imperfect hardware output, a second question is what level of approximation in a tensor-network simulation matches the accuracy of the imperfect quantum computer? The level of approximation in tensor networks, encoded in a parameter $\chi\in[1,2^{N/2}]$ ,controls their run-time: the larger $\chi$, the less approximation is performed, and the longer the simulation takes. For instance, the algorithmic complexity of matrix product states goes as $O(\chi^3)$. Then, one can answer the final question: for a matched output accuracy, which of the approximate tensor-network simulator or quantum computer is faster? This will provide a basis for a hardware-based quantum utility~\cite{PRXQuantum.3.040339,PhysRevA.106.022423,Sreedhar2022,PRXQuantum.4.020304}.

\subsection{Beyond random \texorpdfstring{$3$}{3}-regular graph maximum-cut problems and challenges for a quantum optimization advantage}

Many challenges faced by the proposed quantum-preconditioning method to deliver quantum utility are common to quantum optimization and are active areas of research. We discuss some of the main ones in the following.

\subsubsection{Compilation and efficient swap network for sparse graphs}

Superconducting quantum hardware has a given topology, with a fixed connection between the qubits. This requires the use of two-qubit $\texttt{SWAP}$ gates for achieving quantum logical operations between arbitrary non-neighboring qubits. Compiling QAOA circuits for sparse graph problems using a default swap network (i.e., on a linear arrangement of qubits~\cite{OGorman2019,Weidenfeller2022scalingofquantum}) is far from efficient in terms of circuit depth and number of two-qubit gates. This sets fidelity requirements much higher than what is actually needed under the right compilation strategy. An efficient swap network~\cite{MATSUO20232022EAP1159,Sack2023} for light-cone-induced subgraphs on random $3$-regular graph maximum-cut problems can reduce the number of $\texttt{ISWAP}$ gates by about $52\%$, $73\%$, and, $85\%$ at $p=1$, $p=2$, and $p=3$, respectively~\cite{Dupont2024}. Such smaller circuits would lead to much higher overall output fidelities. Therefore, the sampling requirements for reliably estimating expectation values would also be much lower---in addition to shorter circuits, reducing the execution time---thus greatly speeding up the preconditioning.

For instance, a light-cone-induced subgraph has at most $N=43$ vertices at $p=3$ for a $3$-regular graph. Following Sec.~\ref{sec:noisy_perf}, this translates into about $n_\textrm{2Q}\sim O(10^4)$ via a default swap network or $n_\textrm{2Q}\sim O(10^3)$ with an efficient compilation~\cite{MATSUO20232022EAP1159,Sack2023,Dupont2024}. Assuming an average two-qubit gate fidelity $f=99.5\%$, the default strategy yields an overall fidelity of $F\simeq 10^{-22}$, and $F\simeq 10^{-2}$ for the efficient one. This highlights the importance of the compilation step on noisy quantum devices.

However, finding an efficient swap network is a difficult task relating to graph isomorphism: solving graph isomorphism using a Boolean satisfiability (SAT) formulation is itself an NP-complete problem and scales exponentially with the size of the graph~\cite{MATSUO20232022EAP1159,Sack2023,Dupont2024}. This calls for the development of computationally cheap and scalable approximate methods for encoding graphs onto the hardware-native topology of a quantum computer. In addition, while the focus is on the QAOA, it is possible that other variational quantum circuits could perform better.

\subsubsection{Decomposition technique}

We used the light-cone decomposition technique for classical emulations of QAOA circuits for sparse large-$N$ problems, such as random $3$-regular graphs and the grid energy problem (Sec.~\ref{sec:light_cone_3reg}). At shallow circuit depth, this trades a single $N$-qubit QAOA circuit for $O(N)$ QAOA circuits with $O(\exp p)$ qubits each. While this can be advantageous for small values of $p$, it requires the execution of many subcircuits. For instance, the preconditioning time via state-vector emulations of the grid energy problem (Fig.~\ref{fig:mpes_preconditioning}) took $0.00270(2)$~s at $p=1$ and $32.9(2)$~s at $p=2$ on a 64GB MacBook Pro with an Apple M1 Max chip. The $p=1$ and $p=2$ preconditioning times are within and out of budget, respectively; however, they do not account for the variational search of optimal QAOA angles. A long preconditioning time could be mitigated by parallelizing the execution of subcircuits on quantum hardware if more qubits than necessary are available (e.g., about twenty $50$-qubit circuits can be executed at a time on a $1,000$-qubit device).

Moreover, the light-cone decomposition technique is only applicable to sparse graphs. Dense graphs, such as the Sherrington-Kirkpatrick spin glasses considered, cannot benefit from it. Because running the QAOA on large, dense graph problems is out of scope for noisy devices due to the large number of qubits and gate requirements, this raises the question of whether other decomposition techniques could be envisioned in the context of quantum preconditioning~\cite{Liu2022,patti2022variational,Bechtold_2023,Ponce2023,10363584,Sciorilli2024,Bach2024,Moondra2024,Sundar2024,Maciejewski2024,Acharya2024}.

\subsubsection{Variational optimization}

Another difficulty is the variational optimization of the QAOA or other quantum circuits used for preconditioning. Finding optimal angles is a known challenge for variational optimization due to Barren plateaus~\cite{PhysRevLett.127.120502,Larocca2024}, whereas preconditioning has to happen fast to be within an advantage window.

For the random $3$-regular graph maximum-cut problems and Sherrington-Kirkpatrick spin glasses, we employed tabulated near-optimal parameters~\cite{PhysRevA.103.042612,Wurtz2021,Farhi2022}, thus skipping altogether the variational optimization; however, the uniqueness of the grid energy optimization problem required a variational optimization of the circuit parameters. We note that classes of optimization problems have been found to share near-optimal angles and that such a strategy of transferring parameters from one problem to another could be leveraged~\cite{8916288,PhysRevX.10.021067,PhysRevA.103.042612,Wurtz2021,Galda2021,Farhi2022,basso_et_al:LIPIcs.TQC.2022.7,Galda2023,Jing2023,10.1145/3678184,10.1145/3584706,shaydulin2024evidence}. Other strategies have also been advanced based on interpolation from lower $p$ values and quantum annealing scheduling~\cite{PhysRevX.10.021067,Sureshbabu2024parametersettingin,He2024}; however, this raises the question of the performance of the quantum preconditioner based on nonoptimal angles.

\section{Conclusion}
\label{sec:conclusion}

We have introduced a quantum-preconditioning method for quadratic unconstrained binary optimization problems. Here, preconditioning refers to the task of transforming the input problem into a more suitable form for a solver by improving its convergence. The proposed quantum preconditioner is based the QAOA~\cite{Farhi2014,BLEKOS20241}. We demonstrated that state-of-the-art classical solvers such as simulated annealing and the Burer-Monteiro method benefited from working with the preconditioned input instead of the original one on a range of problems, including Sherrington-Kirkpatrick spin glasses, random $3$-regular graph maximum cut, and a real-world grid energy problem. The amount of preconditioning is controlled by the quantum circuit depth and renders the problem trivial for a classical solver in the infinite-depth limit. We showed that an increased circuit depth for shallow circuits can improve the convergence. Because classical optimization heuristics operate on an accuracy versus run-time basis, any additional time taken for preconditioning a problem should be given to the purely classical solution. Accounting for this, we showed that quantum preconditioning translates into a practical quantum-inspired advantage for random 3-regular graph maximum-cut problems through classical state-vector emulations of the quantum circuits. Finally, we discussed the challenges and prospects for delivering a hardware-based quantum utility in optimization via quantum preconditioning.

The results call for developing a better understanding of what makes a good quantum preconditioner. We found that quantum preconditioning via the QAOA reduces the problem's frustration, making local moves more efficient in approaching the global optimum. In this work, the use of the QAOA as a preconditioner was heuristic, and it is possible that other quantum circuits might be better suited.

It would be interesting to investigate quantum preconditioning in the context of constrained problems---here, we only considered unconstrained problems. Hard constraints are a part of many optimization problems; however, there is a lack of methods for efficiently handling them on contemporary, noisy, quantum computers. Constraints can be encoded in quantum optimization in several ways, and each has its unique challenges. For example, constraints are often encoded as penalty terms~\cite{lucas2014}, but bit strings may violate the penalties. In some cases, special mixers are used to keep the state within the constraint-satisfying subspace of the Hilbert space~\cite{hadfield2019quantum,fuchs2022constraint}, or the problem is encoded such that the Hilbert space contains only constraint-satisfying solutions~\cite{glos2022space,schnaus2024efficient,tabi2020quantum,fuchs2021efficient,hadfield2021representation,fakhimi2021quantum}. In the former (latter) case, the mixer (phase separator) requires multiqubit gates. In our approach, one could quantum precondition a problem in any formulation of the constraints while letting a classical solver handle the hard constraints.

Improving the fidelity of noisy hardware remains the backbone for quantum computers to potentially deliver a quantum advantage in the absence of quantum error correction. While it is common to leverage error-mitigation techniques to enhance noisy expectation values~\cite{RevModPhys.95.045005,Kim2023,RobledoMoreno2024}, one should be mindful of the added time overhead of some of these techniques in a fair benchmark against classical optimization solvers. Error-mitigation methods such as randomized compilation~\cite{PhysRevA.94.052325} or readout-error mitigation~\cite{arrasmith2023development,chen2019detector, maciejewski2020mitigation,nachman2020unfolding,geller2021conditionally,hamilton2020scalable,geller2021toward,bravyi2021mitigating,seo2021mitigation,Karalekas_2020,smith2021qubit,van2022model} may provide a benefit with little overhead. Furthermore, exploring quantum preconditioning or the QAOA~\cite{He2024qec} within the framework of quantum error correction would also be interesting, especially with the advent of fault-tolerant demonstrations~\cite{Bluvstein2024,GoogleQEC,Reichardt2024}. In particular, quantum error correction induces a run-time overhead, which then raises the question of whether there might be an opportunity for a quantum advantage for quantum preconditioning in a fault-tolerant era.

\begin{acknowledgments}
    We gratefully acknowledge B. Evert, M.J. Hodson, S. Jeffrey, F. B. Maciejewski, and D. Venturelli for discussions, input, and collaborations on related works. This work is supported by the U.S. Department of Energy, Office of Science, National Quantum Information Science Research Centers, Superconducting Quantum Materials and Systems Center (SQMS) under Contract No. DEAC02-07CH11359. This research used resources of the National Energy Research Scientific Computing Center, a DOE Office of Science User Facility supported by the Office of Science of the U.S. Department of Energy under Contract No. DE-AC02-05CH11231 using NERSC awards ASCR-ERCAP0028951 and ASCR-ERCAP0031818. The experiments were performed through Rigetti Computing Inc.'s Quantum Cloud Services QCS\textsuperscript{TM} on the Ankaa\textsuperscript{TM}-3 superconducting quantum processor developed, fabricated, and operated by Rigetti Computing Inc.
\end{acknowledgments}

\section*{Author contributions}

M.D. conceived and led the project with support from B.S. M.D. and T.O. performed simulations, data collection, and data analyses. M.D. wrote the manuscript with input from B.S. All coauthors contributed to the discussions leading to the completion of this project.

\section*{Competing interests}

M.D. and B.S. are, have been, or may in the future be participants in incentive stock plans at Rigetti Computing Inc. M.D. and B.S. are inventors on two pending patent applications related to this work (No. 63/631,643 and No. PCT/US2024/033445). The other authors declare that they have no competing interests.

\section*{Data availability}

The problem instance ``\textit{South Carolina 500-Bus System: ACTIVSg500}'' formatted as a MEPS problem is publicly available at~\href{https://doi.org/10.5281/zenodo.14921060}{doi.org/10.5281/zenodo.14921060}~\cite{data_availability}.

\appendix

\section{Problem Classes and Instances}
\label{app:problem_instances}

\subsection{Random \texorpdfstring{$3$}{3}-regular graph maximum-cut problems}

Random $3$-regular graph maximum-cut problems are a standard testbed for investigating the performance of quantum optimization~\cite{Farhi2014,Harrigan2021,Wurtz2021,PhysRevA.103.042612,10313920,PhysRevX.13.041052,PhysRevX.13.041057,PhysRevA.109.012429,Tate2024,Dupont2024,Sachdeva2024,McGeoch2024,Morris2024,Augustino2024,Zhong2024}. They have $N$ vertices, each connected by an edge to three other vertices at random. The edges carry unit weight. These graphs are sparse, with the adjacency matrix $\mathsf{W}$ containing only three nonzero entries per row or column. Despite their apparently simple topology, it is NP-hard to devise an algorithm to find the maximum cut that can guarantee an approximation ratio of at least $99.7\%$~\cite{10.1007/3-540-48523-6_17}, with the current best-known classical algorithms yielding $\alpha\simeq 93.3\%$~\cite{HALPERIN2004169}. We generated random $3$-regular graphs using the Python package NetworkX~\cite{SciPyProceedings_11,STEGER_WORMALD_1999,10.1145/780542.780576}.

The heuristics considered in this work have the potential to deliver a solution with a higher approximation ratio, but they cannot guarantee it \textit{a priori}. Random $3$-regular graph maximum-cut problems have been independently investigated on a range of solvers such as BM~\cite{Lykov2023}, extremal optimization~\cite{PhysRevE.64.026114}, quantum annealing, and SA~\cite{McGeoch2024,PhysRevA.86.052334,PhysRevLett.114.147203}---interestingly, it was shown that quantum annealing is less efficient than SA~\cite{PhysRevA.86.052334,PhysRevLett.114.147203}. Classical quantum-inspired algorithms based on simulated coherent Ising machines~\cite{Goto2016} and simulated bifurcation~\cite{goto2021high,goto2019combinatorial} have also been considered~\cite{Zeng2024}. Random $3$-regular graph maximum-cut problems have also served as a testbed for combinatorial optimization solvers based on graph neural networks~\cite{Schuetz2022,10.1117/12.2529608}, although their performance is far from that of state-of-the-art methods~\cite{Angelini2023,Boettcher2023,Schuetz2023}. We note that a one-to-one benchmark with the existing literature can be difficult as it has focused on the time-to-solution (TTS) metric~\cite{10.3389/fphy.2019.00048}, which relates to the probability of finding the optimal solution. In this work, we measure the performance through the average approximation ratio instead, as quantum preconditioning cannot guarantee the same optimal solution between the original and preconditioned problems, except in the asymptotic $p\to+\infty$ limit. Furthermore, it might be interesting to investigate quantum preconditioning in the context of state-of-the-art classical solvers other than the BM solver and SA considered in this work, such as extremal optimization~\cite{PhysRevE.64.026114} and quantum-inspired algorithms~\cite{Goto2016,10.3389/fphy.2019.00048,goto2021high,goto2019combinatorial,Zeng2024}.

\subsection{Sherrington-Kirkpatrick spin glasses}

SK models~\cite{PhysRevLett.35.1792} correspond to complete graphs, where the edge weights are independent and identically distributed random variables drawn from a normal distribution of zero mean and unit width $\mathsf{W}_{ij}\sim\mathcal{N}(0,1)$ with $\mathsf{W}_{ij}=\mathsf{W}_{ji}$. As a result, the adjacency matrix $\mathsf{W}$ encoding the problem belongs to the Gaussian orthogonal ensemble. Such graphs present two stark differences with $3$-regular graphs: first, they are as dense as possible because of the all-to-all connectivity between vertices; second, they are weighted graphs.

It is NP-hard to find the optimal solution to an SK spin glass; however, an approximate solution arbitrarily close to the optimal one can be obtained in polynomial time through a recently introduced approximate message-passing algorithm~\cite{8948630}: a solution to an $N$-variable problem with an approximation ratio of at least $\alpha=(1-\varepsilon)$ for $\varepsilon>0$ can be found in $O(P_{\varepsilon}N^2)$ with $P_\varepsilon$ a polynomial in $\varepsilon^{-1}$. Nevertheless, SK problems remain highly relevant for developing an understanding thanks to their long-standing connection with statistical physics and benchmarking optimization solvers~\cite{dupont2023quantum,PhysRevA.109.012429,Sundar2024,Farhi2022,PhysRevApplied.22.044074,Maciejewski2024_ndar,Maciejewski2024}.

\subsection{Power-grid optimization problem}

\begin{table}[!t]
    \centering
    \begin{tabular}{cclclcl}
        \hline\hline\\[-0.8em]
        \makecell[c]{\textbf{Number of}\\\textbf{QAOA layers}} & \makecell[c]{\qquad} & \makecell{\textbf{Optimized}\\\textbf{QAOA angles}} & \makecell[c]{\qquad} & \makecell{\textbf{Objective}\\\textbf{value}} & \makecell[c]{\qquad} & \makecell{\textbf{Number of}\\\textbf{nonzero terms}}\\
        \hline\\[-0.6em]
        \makecell[c]{Original} & \makecell[c]{} & \makecell[c]{---} & \makecell[c]{} & \makecell[c]{---} & \makecell[c]{} & \makecell[c]{$226$}\\[0.5em]
        \hline\\[-0.6em]
        \makecell[c]{$p=1$} & \makecell[c]{} & \makecell[c]{$\gamma_1\simeq-0.008418$\\$\beta_1\simeq 2.757259$} & \makecell[c]{} & \makecell[c]{$\simeq 59,721.27$} & \makecell[c]{} & \makecell[c]{$600$}\\[0.5em]
        \hline\\[-0.6em]
        \makecell[c]{$p=2$} & \makecell[c]{} & \makecell[c]{$\gamma_1\simeq-0.006541$\\$\beta_1\simeq 1.074866$\\$\gamma_2\simeq-0.012537$\\$\beta_2\simeq 1.307313$} & \makecell[c]{} & \makecell[c]{$\simeq 60,202.88$} & \makecell[c]{} & \makecell[c]{$1,625$}\\[0.5em]
        \hline\hline\\[-0.8em]
    \end{tabular}
    \caption{Information regarding the MPES problem considered. The QAOA angles were found using the Broyden-Fletcher-Goldfarb-Shanno (BFGS)~\cite{BFGS1,BFGS2,BFGS3,BFGS4} algorithm. The number of nonzero terms correspond to the number of nonzero entries in the correlation matrix of Eq.~\eqref{eq:correlation_matrix}.}
    \label{tab:mpes_qaoa_opt}
\end{table}

Quantum preconditioning on a real-world power-grid optimization problem seeking to compute the MPES of an energy network~\cite{Jing2023,9813820,9948058,Bauer2024} was investigated in Sec.~\ref{sec:grid_opt_problem}. We considered a realistic $500$-bus power system dataset ``\textit{ACTIVSg500}'' mimicking the energy grid of the state of South Carolina in the United States of America~\cite{ACTIVSg500}. The dataset was designed as part of \textit{ARPA-E's GRID DATA} program to be similar to the actual electric grid while containing no confidential information about critical energy infrastructure~\cite{7459256,7515182,7725528,xu2017creation,xu2017}. In Tab.~\ref{tab:mpes_qaoa_opt}, we present information regarding the QAOA settings that were used for preconditioning the $N=180$-variable problem. The reported QAOA angles correspond to minimizing the objective function $\tilde{C}_\textrm{MPES}(\boldsymbol{z})=\frac{1}{2}\sum_{i,j=1}^N\mathsf{W}_{ij}z_iz_j$. The reported objective value is that of the underlying maximum-cut problem of Eq.~\eqref{eq:obj_function_mpes}.

\section{Quantum Preconditioning Eases Problems Hardness}
\label{app:ease_hardness}

\begin{figure}[!t]
    \centering
    \includegraphics[width=1\columnwidth]{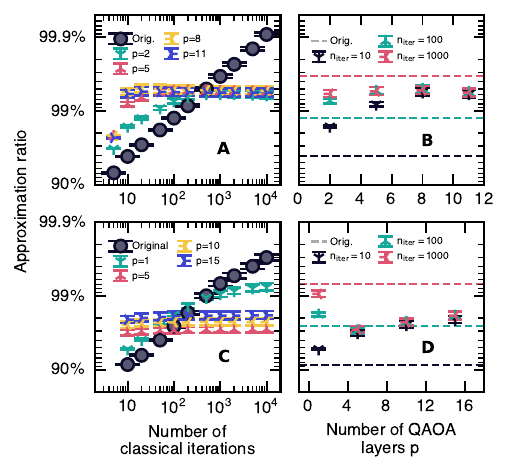}
    \caption{(A) (B) $N=16$ random $3$-regular graph maximum-cut problems. (C) (D) $N=16$ SK spin glasses. (A) (C) Average approximation ratio as a function of the number of classical iterations $n_\textrm{iter}$ for simulated annealing on various, based on quantum preconditioning, for various values of $p$. (B) (D) Average approximation ratio as a function of the number of QAOA layers $p$ used for quantum preconditioning for various number of iterations $n_\textrm{iter}$ in simulated annealing. Each data point is averaged over $200$ randomly generated problem instances. Error bars indicate the standard error of the mean.}
    \label{fig:smallN_precond_sa}
\end{figure}

\begin{figure}[!t]
    \centering
    \includegraphics[width=1\columnwidth]{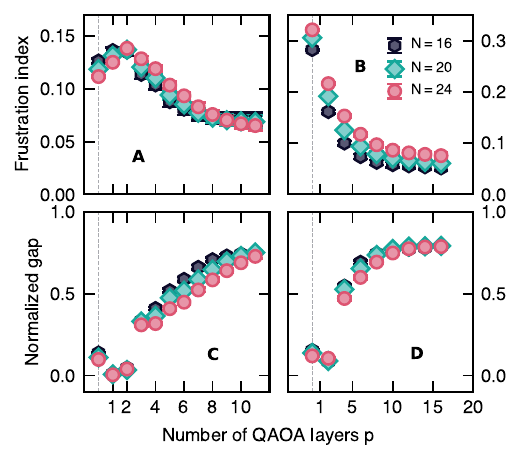}
    \caption{(A) (C) $N$-variable random $3$-regular graph maximum-cut problems. (B) (D) $N$-variable SK spin glasses. (A) (B) Average frustration index defined in Eq.~\eqref{eq:frustration} as a function of the number of QAOA layers $p$. (C) (D) Average normalized gap defined in Eq.~\eqref{eq:gap} as a function of the number of QAOA layers $p$. Vertical dashed lines indicate the quantities evaluated on the original problem $\mathsf{W}$. Each data point is averaged over $200$ randomly generated problem instances. Error bars indicate the standard error of the mean.}
    \label{fig:frustration}
\end{figure}

\subsection{Large-depth quantum preconditioning at small scale}

The focus of the main text was on large problems at shallow circuit depth $p$. We now turn our attention to quantum preconditioning of small-scale ($N=16$) random $3$-regular graph maximum-cut problems and SK spin glasses for large values of $p$. We access large $p$ values via classical emulations using tabulated near-optimal QAOA angles up to $p\leq 11$ for random $3$-regular graph maximum-cut problems~\cite{Wurtz2021,PhysRevA.103.042612} and $p\leq 17$ for SK spin glasses~\cite{Farhi2022,basso_et_al:LIPIcs.TQC.2022.7}. We consider simulated annealing as a solver.

In Fig.~\ref{fig:smallN_precond_sa}, we plot the average approximation ratio as a function of the number of iterations in simulated annealing for a fixed value of $p$ and as a function of $p$ for a fixed number of iterations. We observe that increasing $p$ in quantum preconditioning makes simulated annealing converge to an asymptotic approximation ratio value faster, i.e., in fewer iterations.

For a large number of iterations, we find that the approximation saturates to a value $\alpha < 1$, as anticipated, because the original and preconditioned problems do not share the optimal solution except in the asymptotic $p\to+\infty$ limit. Nevertheless, we find that increasing $p$ increases the asymptotic approximation ratio value for SK spin glasses. It is less clear for the random $3$-regular graph maximum-cut problems (Fig.~\ref{fig:smallN_precond_sa}a). We note that the random $3$-regular graph maximum-cut problems are more likely to have degenerate solutions than the SK spin glasses beyond the global $\mathbb{Z}_2$ spin flip symmetry, due to the nature of their weights $\mathsf{W}_{ij}$: unity versus normally distributed. As such, the correlation matrix $\mathsf{Z}^{(p)}$ may not tend to the desired asymptotic $p\to+\infty$ limit discussed in Sec.~\ref{sec:quantum_precond}. Another preconditioner may be more suited at large $p$.

\subsection{Solver-agnostic quantities}

We consider solver-agnostic quantities related to the problem's hardness for random $3$-regular graph maximum-cut problems and SK spin glasses. First, we compute the frustration index~\cite{10.1093/acprof:oso/9780198570837.001.0001,https://doi.org/10.1002/net.21907,JMLR:v21:19-368}
\begin{equation}
   f = \frac{1}{2}+\frac{\sum\nolimits_{i,j=1}^N\mathsf{W}_{ij}z_{\textrm{opt},i}z_{\textrm{opt},j}}{2\sum\nolimits_{i,j=1}^N\vert\mathsf{W}_{ij}\vert};\quad\mathsf{W}_{ij}\leftrightarrow\mathsf{Z}^{(p)}_{ij},
   \label{eq:frustration}
\end{equation}
where $f\in[0,1]$ measures how strongly the optimal solution $\boldsymbol{z}_{\textbf{opt}}$ is frustrated with respect to an input problem $\mathsf{W}$ or $\mathsf{Z}^{(p)}$. Second, we compute the normalized gap
\begin{equation}
   \Delta = \bigl(\sigma_2 - \sigma_1\bigr)\bigr/\bigl(\sigma_N - \sigma_1\bigr),
   \label{eq:gap}
\end{equation}
where $\sigma_{i=1,2,\ldots N}\geq 0$ are the singular values sorted in ascending order of an input problem $\mathsf{I} + \mathsf{W}$ or $\mathsf{I} + \mathsf{Z}^{(p)}$. The normalized gap $\Delta\in[0,1]$ indicates how accurately the matrix can be approximated by a rank-$1$ matrix. Indeed, in the asymptotic $p\to+\infty$ limit in the presence of a doubly degenerate optimal solution $\pm\boldsymbol{z}_{\textbf{opt}}$, one has $\mathsf{I} + \mathsf{Z}^{(\infty)}=\boldsymbol{z}_{\textbf{opt}}\boldsymbol{z}_{\textbf{opt}}^T$, which is a rank-$1$ matrix, where in that case, the normalized gap is $\Delta=1$.

\begin{figure}[!t]
    \centering
    \includegraphics[width=1\columnwidth]{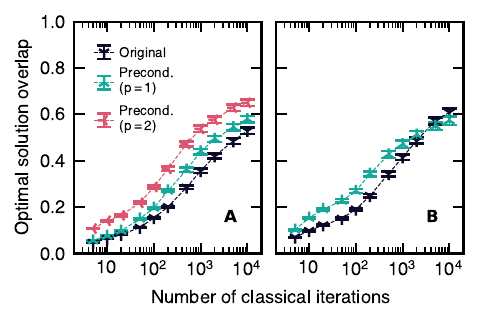}
    \caption{Average optimal solution overlap as defined in Eq.~\eqref{eq:zopt_overlap} as a function of the number of classical iterations in simulated annealing. (A) $N=128$-variable random $3$-regular graph maximum-cut problems. (B) $N=128$-variable SK spin glasses. Each data point is averaged over $200$ randomly generated problem instances. Error bars indicate the standard error of the mean.}
    \label{fig:overlap_sa_zopt}
\end{figure}

In Fig.~\ref{fig:frustration}, we plot these quantities on both the original problem $\mathsf{W}$ and the preconditioned ones $\mathsf{Z}^{(p)}$. We find that for the largest $p$ values, the frustration index and gap asymptotically reach their $p\to+\infty$ value, i.e., $f\to 0$ and $\Delta\to 1$. This highlights the potential of quantum preconditioning to ease the hardness of challenging optimization problems. At a shallow circuit depth $p$, such as investigated in the rest of this work, the behavior of $f$ and $\Delta$ is much less clear. In particular, for random $3$-regular graph maximum-cut problems, we observe an increase in frustration with $p$ at small $p$ in Fig.~\ref{fig:frustration}a. This is counterintuitive, as the quantum-preconditioned problem is actually easier to solve for algorithms such as simulated annealing (Fig.~\ref{fig:sa_preconditioning}) and Burer-Monteiro (Fig.~\ref{fig:burer2002_preconditioning}; however, we observe a change of slope for $p=2$, which indicates a drop in frustration for $p\geq 3$, which may further facilitate convergence at larger circuit depths for solvers relying on quantum preconditioning. Testing this hypothesis is out of scope for this work given the challenges of scaling up quantum circuits to $p=3$ and beyond on relevant large problem sizes $N$ (see discussions in Sec.~\ref{sec:propsects_3reg}).

\subsection{Optimal Solution Overlap}

Next, we consider the overlap of a sampled solution $\boldsymbol{z}$ via simulated annealing with the optimal one $\boldsymbol{z}_\textbf{opt}$. This is defined as
\begin{equation}
    q^2 = \left(\frac{1}{N}\sum\nolimits_{i=1}^Nz_{\textrm{opt},i}z_i\right)^2,
    \label{eq:zopt_overlap}
\end{equation}
reminiscent of the Edwards-Anderson overlap~\cite{SFEdwards_1975,SFEdwards_1976,PhysRevLett.50.1946}. In Fig.~\ref{fig:overlap_sa_zopt}, we plot the overlap with the optimal solution as a function of the number of classical iterations in simulated annealing for $N=128$ random $3$-regular graph maximum-cut problems and SK spin glasses.

\begin{figure}[!t]
    \centering
    \includegraphics[width=1\columnwidth]{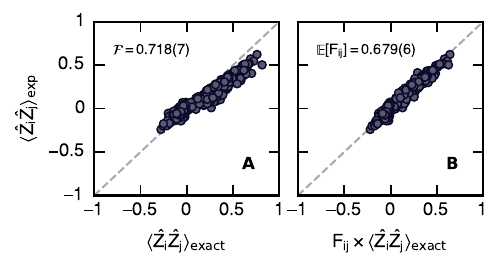}
    \caption{There are $600$ data points corresponding to $p=1$ QAOA circuits of the South Carolina grid energy problem. (A) Experimental two-point correlation versus the exact one. A least-squares fitting of the form $y=\mathcal{F}x$ yields $\mathcal{F}=0.718(7)$. (B) Experimental two-point correlation versus the exact one rescaled by the corresponding circuit fidelity $F_{ij}$. The average circuit fidelity across all the circuits is $\mathbb{E}[F_{ij}]=0.679(6)$.}
    \label{fig:meps_zizj_experiment}
\end{figure}

In both cases, we find that for a fixed number of iterations of the classical solver, quantum preconditioning yields solutions $\boldsymbol{z}$ with a greater overlap with the optimal solution than simulated annealing working on the original problem. The overlap follows the trend of the approximation ratio in Figs.~\ref{fig:sa_3reg_preconditioning}c and.~\ref{fig:sk_N_preconditioning}c, where we anticipate that the original problem will asymptotically reach $q^2\to 1$, while the preconditioned problems, not sharing the same optimal solutions, will result in $q^2<1$. The larger overlap supports quantum preconditioning easing the convergence of classical solvers toward near-optimal solutions.

\section{On the Light-Cone Decomposition Technique}
\label{app:lightcone_technique}

On sparse problem instances, such as random $3$-regular graphs and the South Carolina grid network, one can leverage a light-cone decomposition technique for estimating expectation values such as $\langle\hat{Z}_i\hat{Z}_j\rangle$ for preconditioning the problem.

Indeed, a $p$-layer QAOA circuit as defined in Eq.~\eqref{eq:qaoa} induces quantum coherence around each qubit and its $p$-nearest neighbors, as per the underlying graph topology defined by $\mathcal{G}$. This defines a light-cone-induced subgraph $\mathcal{G}_i$ for each of the qubits or problem variables $i$, which can straightforwardly be obtained. For computing an expectation value $\langle\hat{Z}_i\hat{Z}_j\cdots\hat{Z}_m\rangle$, one may only consider the union of the light-cone-induced subgraphs for each of the involved variables, i.e., $\mathcal{G}_{ij\cdots m}=\mathcal{G}_i\cup\mathcal{G}_j\cdots\cup\mathcal{G}_m$. Executing the QAOA on either $\mathcal{G}$ or $\mathcal{G}_{ij\cdots m}$ will lead to the same expectation value $\langle\hat{Z}_i\hat{Z}_j\cdots\hat{Z}_m\rangle$.

The light-cone technique presents an advantage for computing $\langle\hat{Z}_i\hat{Z}_j\cdots\hat{Z}_m\rangle$ when $\mathcal{G}_{ij\cdots m}$ is smaller than $\mathcal{G}$, as the resulting QAOA execution involves fewer qubits and gates. The price is that of a different QAOA execution for every single desired expectation value $\langle\hat{Z}_i\hat{Z}_j\cdots\hat{Z}_m\rangle$ corresponding to different subgraphs $\mathcal{G}_{ij\cdots m}$.

For a graph with $N$ nodes with a maximum degree $d$, the number of nodes in a light-cone-induced subgraph by the QAOA with depth $p$ for computing a two-point correlation $\langle\hat{Z}_i\hat{Z}_j\rangle$ is \textit{at most}
\begin{equation}
    \textrm{min}\left(1+\frac{2d\left[(d-1)^p - 1\right]}{d-2}, N\right).
\end{equation}
For example, this upper bound is saturated for $d$-regular graphs such as the random $3$-regular graph considered. For $N$-variable all-to-all graphs where $d=N-1$, such as the SK problems considered, the light-cone subgraphs involve all the $N$ nodes, meaning that the execution of the QAOA on the full graph is required. This worst-case bound is not saturated for the grid energy network. The light-cone technique is exact and leads to smaller simulations for sparse graphs at shallow QAOA depth $p$ while mapping to the standard QAOA for dense graphs or large $p$ values.

\section{Experimental Implementation of the South Carolina Maximum Power Exchange Section Problem}
\label{app:experiments_mpes}

We implement quantum preconditioning at $p=1$ on the Rigetti Ankaa-3 superconducting quantum computer for the South Carolina grid energy problem of Sec.~\ref{sec:grid_opt_problem}. Using the light-cone decomposition technique for computing necessary expectation values, this requires executing $600$ QAOA circuits ranging from $N=4$ to $N=14$ qubits (see Tab.~\ref{tab:mpes_qaoa_opt}). We use the transpilation strategy of Sec.~\ref{sec:sampling_time} in terms of one- and two-qubit gates. The average fidelity of the two-qubit $\texttt{ISWAP}$ gates is approximately $98.6\%$, of the one-qubit gate $\texttt{RX}$ is approximately $99.8\%$, and that of the readout is approximately $96.7\%$~\cite{PhysRevA.77.012307}. Computations are performed on a linear chain of qubits with the phase separator of the QAOA compiled via an efficient swap network~\cite{MATSUO20232022EAP1159,Sack2023}. Through a na\"ive swap network, $N$-qubit QAOA circuits at $p=1$ require $3N^2/2$ two-qubit $\texttt{ISWAP}$ gates~\cite{OGorman2019}---i.e., $24$ for $N=4$ and $294$ for $N=14$. Instead, the efficient swap network strategy makes this number range between $6$ and $83$ for the $600$ circuits at hand---this is a three-to-four-fold improvement. We collect $K=10^4$ bit strings for each circuit for estimating expectation values.

In Fig.~\ref{fig:meps_zizj_experiment}a, we show the experimental two-point correlation versus the exact one obtained via classical state-vector emulations. As discussed in Sec.~\ref{sec:noisy_perf}, we anticipate a simple depolarizing noise model to rescale expectation values by a factor corresponding to the circuit fidelity. We find that the data roughly follows a linear relationship with a slope $\mathcal{F}=0.718(7)$ obtained by least-squares fitting. In Fig.~\ref{fig:meps_zizj_experiment}b, we rescale the exact value of each correlation by the corresponding circuit fidelity $F_{ij}$, which has been evaluated independently by multiplying the fidelity of all present individual operations, including one-qubit gates, two-qubit gates, and readout. The average fidelity over the $600$ circuits is $\mathbb{E}[F_{ij}]=0.679(6)$, which is relatively close to $\mathcal{F}$ and explains why the rescaled data of Fig.~\ref{fig:meps_zizj_experiment}b follows the $y=x$ line more closely.

\begin{figure}[!t]
    \centering
    \includegraphics[width=1\columnwidth]{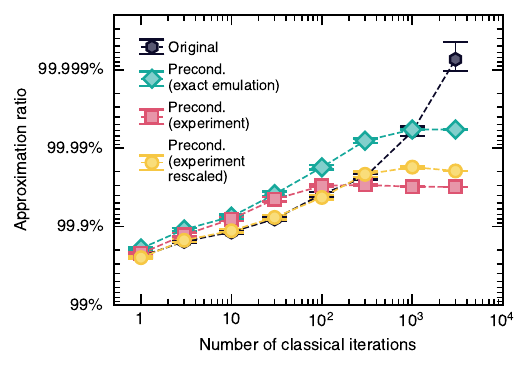}
    \caption{Average approximation ratio against the number of iterations for the BM solver for the South Carolina grid energy problem. Different input problems are considered: the original problem, $p=1$ preconditioned problem via exact circuit emulations; the $p=1$ preconditioned problem via experimental computations; and the $p=1$ preconditioned problem via exact circuit emulations with values rescaled by the circuit fidelity. Each data point is averaged over $100$ random independent runs of the BM solver. Error bars indicate the standard error of the mean.}
    \label{fig:meps_ar_experiment}
\end{figure}

Having preconditioned the problem, we now solve it using the classical BM algorithm---a comparison against SA in Fig.~\ref{fig:mpes_preconditioning} showed that BM outperformed it by one order of magnitude, thus favoring BM as the solver of choice. In Fig.~\ref{fig:meps_ar_experiment}, we plot the approximation ratio versus the number of iterations. We find that the experimentally preconditioned problem provides an advantage when compared to the original problem. We observe that the preconditioned problem rescaled by the circuit fidelity does not yield an advantage here. At current noise levels, the benefit of preconditioning by implementing the QAOA in experiment is lower than that of the exact classical state-vector emulation; however, we emphasize that we are achieving average approximation ratios above $\alpha\simeq 99.97\%$ and that the observed differences are extremely small.

This highlights the experimental applicability of the proposed quantum-preconditioning method, although more work is needed for quantum preconditioning to potentially deliver utility. Understanding how noise affects quantum preconditioning would be valuable.

\section{Implementation of Classical Solvers}
\label{app:classical_solvers}

In this work, we focus on two state-of-the-art classical optimization solvers, which are both heuristics: SA and the BM algorithm. We rely on existing efficient C++ implementations for both solvers.

\subsection{Simulated annealing}

Simulated annealing is a physics-inspired approach to solving that treats an objective function such as Eq.~\eqref{eq:obj_function} as a classical Ising model for which one wants to find the ground state. The search is based on a Metropolis-Hastings Markov-chain Monte Carlo algorithm~\cite{10.1063/1.1699114,10.1093/biomet/57.1.97} with a Boltzmann distribution. It starts at a high temperature, which is reduced throughout the iterations according to a predefined schedule such that one samples low-energy states, and ideally the ground state in the last step.

For an $N$-variable problem, an iteration, also known as a sweep, attempts to flip the sign of $N$ variables uniformly at random. Each flip is accepted (or discarded) according to a Metropolis-Hastings probability. The initial configuration $\boldsymbol{z}$ is generated at random among the $2^N$ possibilities. Each sweep $\ell$ works with a fixed temperature $T_\ell$. We employ a geometric temperature schedule such that for a total of $M$ sweeps (defined by the user), the temperature at iteration $\ell=1,~2,~\ldots,~M$ reads
\begin{equation}
    T^{-1}_\ell=\exp\Biggl(\ln T_\textrm{hot}^{-1} + \ell\frac{\ln T_\textrm{cold}^{-1} - \ln T_\textrm{hot}^{-1}}{M}\Biggr),
    \label{eq:temp_schedule}
\end{equation}
where $T_\textrm{hot}\geq T_\textrm{cold}\in\mathbb{R}$ are the initial and final temperatures, respectively. Finding optimal initial and final temperatures may require its own analysis and fine-tuning. The shape of the temperature schedule can also be optimized beyond a simple geometric descent. Here, we employ a heuristic to set $T_\textrm{hot}$ and $T_\textrm{cold}$ based on the $n$ edges and weights of the input problem. This has an algorithmic complexity of $O(n)$, as visible in Appendix~\ref{app:iterations_to_run-time}. This step, which is only performed once at the beginning, is independent of the number of sweeps and typically not a bottleneck for SA.

First, we define the effective field experienced by a spin $i$ due to two-body interactions with its nearest neighbors
\begin{equation}
    h_i=\sum\nolimits_{j=1}^N\bigl\vert\mathsf{W}_{ij}\bigr\vert.
\end{equation}
In the context of the preconditioned problems, one substitutes $\mathsf{W}_{ij}$ with $\mathsf{Z}_{ij}^{(p)}$ as defined in Eq.~\eqref{eq:correlation_matrix}. Initially, when the temperature is hot, we want fast mixing such that all spins can be flipped with a probability of $50\%$ (number chosen arbitrarily). Thus, $0.5=\exp(-2\max\nolimits_ih_i/T_\textrm{hot})$, where $2\max\nolimits_ih_i$ corresponds to the largest possible energy gap between two configurations after a spin flip. Inverting the relation, one gets
\begin{equation}
    T_\textrm{hot}=2\max\nolimits_ih_i\bigr/\ln 2.
\end{equation}
At the coldest (and final) temperature, we bound the probability of exciting any of the $N$ spins by $1\%$. Assuming that only spins with a minimal energy gap are excitable, one has $0.01=\nu\exp(-2\min\nolimits_{ij}\vert\mathsf{W}_{ij}\vert/T_\textrm{cold})$, where we have approximated the minimal energy gap by $2\min\nolimits_{ij}\vert\mathsf{W}_{ij}\vert$. Inverting the relation, one gets
\begin{equation}
    T_\textrm{cold}=2\min\nolimits_{ij}\vert\mathsf{W}_{ij}\vert\bigr/\ln\bigl(100\nu\bigr),
\end{equation}
where $\nu\in[1, N]$, counts for $i=1,\ldots, N$, the number of values $\min_i\vert\mathsf{W}_{ij}\vert$ equal to $\min\nolimits_{ij}\vert\mathsf{W}_{ij}\vert$, i.e., spins with a minimal energy gap. For instance, for the original unit-weight random $3$-regular graph maximum-cut problems, $h_i=3\forall i$, such that $T_\textrm{hot}=6\ln2$ and $T_\textrm{cold}=6/\ln(100N)$, since $\nu=N$ in this case.

The SA implementation used throughout this work relies on the Python package~\texttt{dwave-samplers}~1.4.0~\cite{dwave_sa}, although the underlying code is in C++ for efficiency. The temperature-setting heuristic described above is the default one in this implementation.

\begin{figure}[!t]
    \centering
    \includegraphics[width=1\columnwidth]{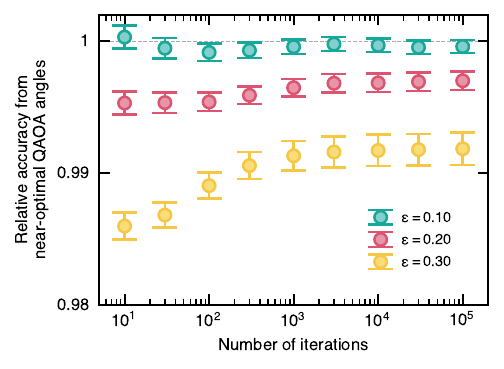}
    \caption{Effect on the approximation ratio returned by simulated annealing versus the number of iterations due to variations in QAOA angles defined by $\varepsilon>0$, relative to the near-optimal ($\varepsilon=0$) case. SK problem instances with $N=64$ variables at $p=1$ are considered, where each data point is averaged over $500$ randomly generated instances. Error bars indicate the standard error of the mean.}
    \label{fig:distance_opt_angles}
\end{figure}

\subsection{Burer-Monteiro solver}

The BM solver eliminates the positive semidefinite constraint of the GW algorithm for the maximum-cut problem. The GW algorithm is based on semidefinite programming methods and the solver guarantees the highest possible approximation ratio $\alpha\simeq 0.878$~\cite{Goemans1995,williamson2011}. Thus, the BM solver involves a nonconvex objective function, which may result in multiple local nonglobal minima. In practice, this is not an issue, and the BM solver is considered one of the best heuristics to date for the maximum-cut problem~\cite{DunningEtAl2018}. Its foundations are still intensively studied by the community; see, e.g.,~\cite{waldspurger2020rank,boumal2020deterministic} and references therein.

We employ the MQLib implementation of the BM solver~\cite{DunningEtAl2018}, written in C++ for efficiency. This implementation does not readily expose any hyperparameters for the solver and we keep the default values. This is why, unlike for SA, the run-time is strictly proportional to the number of iterations (see Appendix~\ref{app:iterations_to_run-time}). Our only modification of the MQLib code exposes the number of iterations performed by the solver in addition to its run-time.

\section{Quantum Preconditioning Performance Away from Near-Optimal QAOA Angles}

We investigate the performance of the proposed quantum-preconditioning approach away from near-optimal QAOA angles. We perform numerical simulations on $N=64$ SK problem instances at $p=1$ with QAOA angles defined as
\begin{equation}
    \gamma=\bigl(\varepsilon\cos\theta + 0.5\bigr)/\sqrt{N},~~\beta=\varepsilon\sin\theta+\pi/8,
\end{equation}
where $\varepsilon$ measures the Euclidean distance to the near-optimal tabulated QAOA angles and $\theta$ is a random number uniformly drawn from the range $[0,2\pi]$. For $\varepsilon=0$, one recovers the near-optimal tabulated angles~\cite{Farhi2022}.

In Fig.~\ref{fig:distance_opt_angles}, we show the effect of $\varepsilon>0$ relative to the $\varepsilon=0$ case on the performance of simulated annealing versus the number of iterations, as measured by $\alpha(\varepsilon)/\alpha(\varepsilon=0)$, where $\alpha$ is the average approximation ratio over problem instances and samples. We observe that the quantum-preconditioning approach is quite robust to variations of QAOA angles. The sensitivity to $\varepsilon$ for larger $p$ remains open.

Finally, we note that for random $3$-regular graphs, the tabulated angles that we used throughout this work are not optimal on an instance-by-instance basis but are only near-optimal on average~\cite{Wurtz2021}, meaning that they correspond to an effective finite $\varepsilon$ value; however, this near-optimality of the angles was enough to provide a quantum-inspired advantage via quantum preconditioning.

\section{Terms in Quantum-Preconditioned Random \texorpdfstring{$3$}{3}-Regular Graph Maximum-Cut Problems}
\label{app:n_nonzero_terms_3reg}

\begin{figure}[!t]
    \centering
    \includegraphics[width=1\columnwidth]{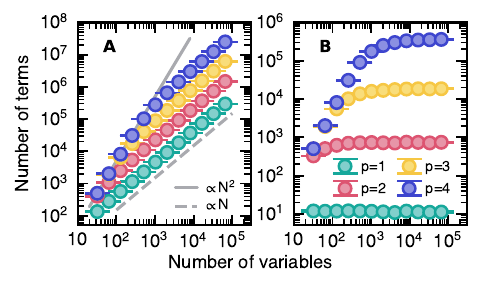}
    \caption{(A) Average number of nonzero terms in quantum-preconditioned random $3$-regular graph maximum-cut problems for various values of $p$ as a function of the number of variables $N$. (B) Average number of terms to precondition in quantum-preconditioned random $3$-regular graph maximum-cut problems for various values of $p$ as a function of the number of variables $N$. It is smaller than or equal to the number of nonzero term via caching the preconditioned value of light-cone-induced tree subgraphs. Each data point is averaged over $200$ randomly generated problem instances. Error bars indicate the standard error of the mean.}
    \label{fig:terms_precond_3reg}
\end{figure}

\begin{table}[!t]
    \centering
    \begin{tabular}{cclll}
        \hline\hline\\[-0.8em]
        \makecell[c]{\textbf{Number of}\\\textbf{QAOA layers}} & \makecell[c]{\qquad\quad} & \makecell{\textbf{Number of}\\\textbf{nonzero terms}\\\textbf{for}~$\boldsymbol{p\ll\ln N}$} & \makecell[c]{\qquad\quad} & \makecell{\textbf{Number of}\\\textbf{terms to precondition}\\\textbf{for}~$\boldsymbol{p\ll\ln N}$}\\
        \hline\\[-0.6em]
        \makecell[c]{Original} & \makecell[c]{} & \makecell[c]{1.5N} & \makecell[c]{} & \makecell[c]{---}\\[0.3em]
        \hline\\[-0.6em]
        \makecell[c]{$p=1$} & \makecell[c]{} & \makecell[c]{$4.500008(8)N$} & \makecell[c]{} & \makecell[c]{$11.2(4)$}\\[0.3em]
        \hline\\[-0.6em]
        \makecell[c]{$p=2$} & \makecell[c]{} & \makecell[c]{$22.5000(2)N$} & \makecell[c]{} & \makecell[c]{$740(17)$}\\[0.3em]
        \hline\\[-0.6em]
        \makecell[c]{$p=3$} & \makecell[c]{} & \makecell[c]{$94.500(2)N$} & \makecell[c]{} & \makecell[c]{$18,581(231)$}\\[0.3em]
        \hline\\[-0.6em]
        \makecell[c]{$p=4$} & \makecell[c]{} & \makecell[c]{$382.45(3)N$} & \makecell[c]{} & \makecell[c]{$360,937(2,832)$}\\[0.3em]
        \hline\hline\\[-0.8em]
    \end{tabular}
    \caption{Asymptotic behaviors from Fig.~\ref{fig:terms_precond_3reg}. Numbers of nonzero terms for $p\ll\ln N$ extracted by a linear fit of the four largest problem sizes. The numbers of terms to precondition for $p\ll\ln N$ correspond to the average value of the largest problem size with $N=2^{16}$ variables.}
    \label{tab:nterms_asymptotic_3reg}
\end{table}

Random $3$-regular graph maximum-cut problems are sparse: the upper triangle of the adjacency matrix $\mathsf{W}$ contains $3N/2$ entries. The QAOA with $p$ layers induces nonzero correlations $\langle\hat{Z}_i\hat{Z}_j\rangle$ at a distance $2p$ between variables $i$ and $j$ via a light cone. This results in additional nonzero terms in the preconditioned problem when compared to the original one. In Fig.~\ref{fig:terms_precond_3reg}a, we show the number of terms $n$ as a function of the number of variables $N$ for various values of $p$. We observe two asymptotic behaviors: $n\sim O(N)$ for $p\ll\ln N$ and $n\sim O(N^2)$ for $p\gg\ln N$, understood through the topology of random $3$-regular graphs having an average distance between two variables going as $\ln N$.

Although the number of nonzero terms grows with $N$, only an $N$-independent number $O(\exp p)$ of terms needs to be individually preconditioned for $p\ll\ln N$. This is because asymptotically with $N$, most of the correlations $\langle\hat{Z}_i\hat{Z}_j\rangle$ are identical: they correspond to light-cone-induced tree subgraphs. A tree is straightforwardly detected as a graph for which the number of edges $n$ is $N=n+1$ and its corresponding correlation value $\langle\hat{Z}_i\hat{Z}_j\rangle$ only needs to be computed once and cached.

In Tab.~\ref{tab:nterms_asymptotic_3reg}, we report asymptotic numerical values. We conjecture that the prefactor in the $O(N)$ scaling for the average number of nonzero terms relates to a sequence of integers $\{3,~9,~45,~189,~765,~\ldots\}$, which is, however, not referenced in the online encyclopedia of integer sequences (OEIS).

\begin{figure*}[!t]
    \centering
    \includegraphics[width=1\textwidth]{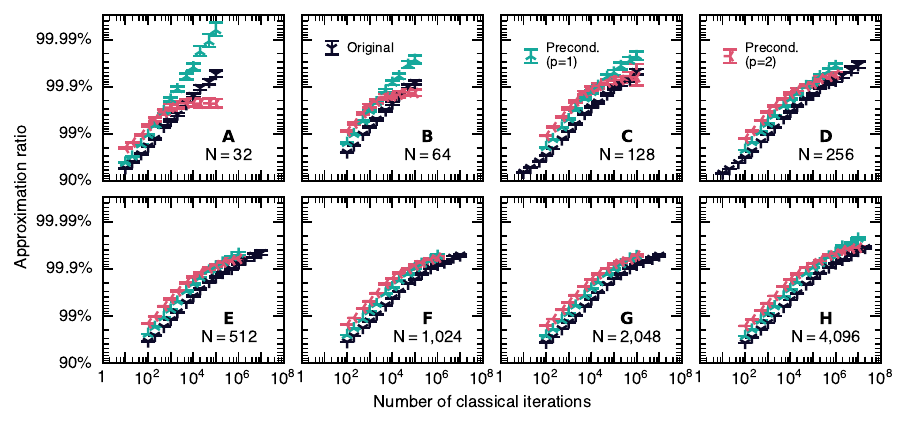}
    \caption{Average performance (approximation ratio versus the number of iterations) for the maximum-cut problem on $N$-variable random $3$-regular graphs via the classical SA solver based on the original and quantum-preconditioned problems. (A) $N=32$ variables. (B) $N=64$ variables. (C) $N=128$ variables. (D) $N=256$ variables. (E) $N=512$ variables. (F) $N=1,024$ variables. (G) $N=2,048$ variables. (H) $N=4,096$ variables. Each data point is averaged over $200$ randomly generated problem instances. Error bars indicate the standard error of the mean.}
    \label{fig:sa_3reg_preconditioning}
\end{figure*}

\section{Semi-Analytical Expression for the Correlation Matrix at \texorpdfstring{$p=1$}{p=1}}
\label{app:corr_formula}

The proposed quantum preconditioning can be performed at $p=1$ for arbitrary $N$-variable problems classically and semianalytically using a back-propagation technique for evaluating expectation values $\langle\hat{Z}_i\hat{Z}_j\rangle_{p=1}$ as a function of the QAOA parameters $\gamma_1\equiv\gamma$ and $\beta_1\equiv\beta$, as defined in Eq.~\eqref{eq:qaoa}. For an arbitrary objective function (arbitrary adjacency matrix $\mathsf{W}$) of the form of Eq.~\eqref{eq:obj_function}, the complexity for computing the full correlation matrix is $O(N^3)$. We simply state the result derived in Ref.~\cite{PhysRevA.109.012429} for completeness:
\begin{align}
    \bigl\langle\hat{Z}_i\hat{Z}_j\bigr\rangle_{p=1}=& -\sin\bigl(2\beta\bigr)\cos\bigl(2\beta\bigr)\sin\bigl(\gamma \mathsf{W}_{ij}\bigr)\nonumber\\
    &\times\Biggl[\prod\nolimits_{k\neq i,j}\cos\bigl(\gamma \mathsf{W}_{ik}\bigr)+\prod\nolimits_{k\neq i,j}\cos\bigl(\gamma \mathsf{W}_{jk}\bigr)\Biggr]\nonumber\\
    &-\frac{\sin^2\bigl(2\beta\bigr)}{2}\Biggl[\prod\nolimits_{k\neq i,j}\cos\gamma \bigl(\mathsf{W}_{ik}+\mathsf{W}_{jk}\bigr)\nonumber\\
    &-\prod\nolimits_{k\neq i,j}\cos\gamma\bigl(\mathsf{W}_{jk}-\mathsf{W}_{ik}\bigr)\Biggr].
    \label{eq:zz_formula_qaoa_p1}
\end{align}
We use Eq.~\eqref{eq:zz_formula_qaoa_p1} for preconditioning the Sherrington-Kirkpatrick spin-glass problems studied in the main text.

\begin{figure*}[!t]
    \centering
    \includegraphics[width=1\textwidth]{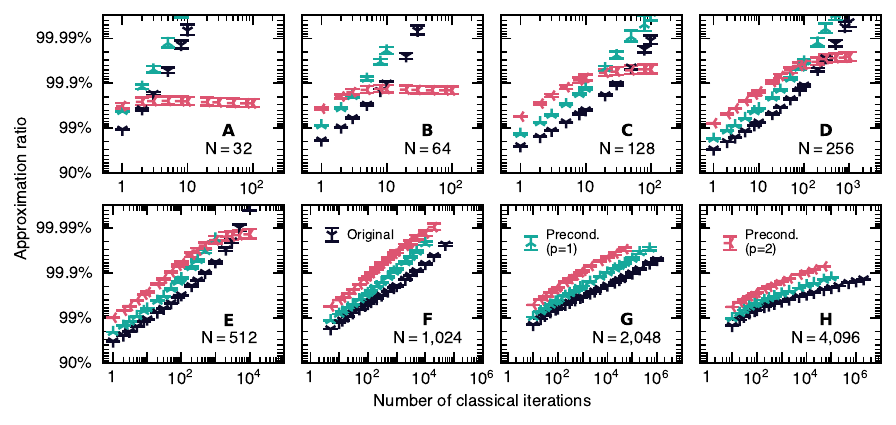}
    \caption{Average performance (approximation ratio versus the number of iterations) for the maximum-cut problem on $N$-variable random $3$-regular graphs via the classical BM solver based on the original and quantum-preconditioned problems. (A) $N=32$ variables. (B) $N=64$ variables. (C) $N=128$ variables. (D) $N=256$ variables. (E) $N=512$ variables. (F) $N=1,024$ variables. (G) $N=2,048$ variables. (H) $N=4,096$ variables. Each data point is averaged over $200$ randomly generated problem instances. Error bars indicate the standard error of the mean.}
    \label{fig:burer2002_3reg_preconditioning}
\end{figure*}

\section{Performance Against Approximate Classical Tensor Network Emulations}
\label{app:perf_approx_mps}

\begin{figure}[!t]
    \centering
    \includegraphics[width=1\columnwidth]{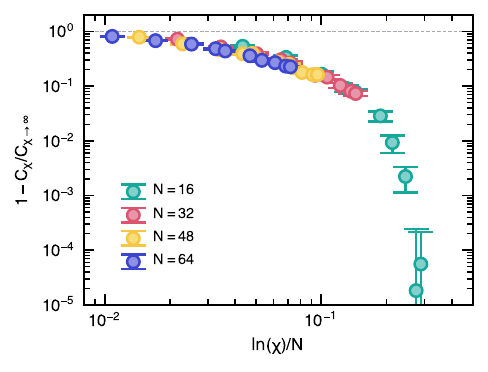}
    \caption{Performance of simulated annealing for quantum-preconditioned SK problems with approximate MPS emulations. Ratio $1-C_{\chi}/C_{\chi\to\infty}$ against $\ln\chi/N$, where $\chi$ is the MPS bond dimension and $N$ is the number of variables. The value $C_{\chi}$ is the average objective value of $10^3$ solutions sampled with simulated annealing. Each data point is averaged over $50$ randomly generated problem instances. Error bars indicate the standard error of the mean.}
    \label{fig:mps_approx}
\end{figure}

To complement the discussion of Sec.~\ref{sec:approx_tn_vs_nisq}, we perform classical matrix-product-state (MPS) emulations of the $p=1$ QAOA used to precondition an ensemble of $N$-variable SK problems. We employ the Python library \textit{quimb}~\cite{gray2018quimb} for the MPS emulations. The QAOA is emulated approximately, with the emulation quality controlled by the MPS bond dimension $\chi$. Then, we use simulated annealing with a fixed number of $10^4$ iterations to evaluate the performance of the classical solver against approximately quantum-preconditioned problems.

Inspired by Ref.~\cite{PRXQuantum.3.040339}, and for each problem instance, we consider the average objective value $C_{\chi}$ over $10^3$ solutions sampled by simulated annealing. The comparison point is with respect to the solutions returned from perfectly quantum-preconditioned problems, i.e., $C_{\chi\to\infty}$, which we can efficiently access at $p=1$ based on Sec.~\ref{app:corr_formula}.

In Fig.~\ref{fig:mps_approx}, we observe a data collapse for $C_{\chi}/C_{\chi\to\infty}$ as a function of the rescaled variable $\ln\chi/N$. This behavior for the output of simulated annealing with quantum preconditioning was also observed for the standard QAOA~\cite{PRXQuantum.3.040339}, on which the quantum-preconditioning step is based.

For example, one needs $\ln\chi/N\gtrsim 0.1$ for the average quality solutions of approximately quantum-preconditioned simulated annealing to be $90\%$ as good as the exactly ($\chi\to +\infty$) preconditioned one.

\section{Performance Versus Number of Iterations}
\label{app:perf_niter}

\subsection{Random \texorpdfstring{$3$}{3}-regular graph maximum-cut problems}

We investigate the performance for the maximum-cut problem on $N$-variable random $3$-regular graphs based on the original and quantum-preconditioned problems. We consider SA and the BM solver on problem sizes ranging from $N=32$ to $N=4,096$. For each problem size, we average the performance over $200$ randomly generated instances. The preconditioning is performed using the light-cone technique described in Sec.~\ref{sec:light_cone_3reg}.

We report data in Figs.~\ref{fig:sa_3reg_preconditioning} and~\ref{fig:burer2002_3reg_preconditioning}. We observe an advantage for the preconditioned problems against the original one as a function of the number of iterations. This advantage is enhanced by going from $p=1$ to $p=2$. We observe that for a large number of iterations, the approximation ratio for the quantum-preconditioned problem at $p=2$ saturates. This is because the original and preconditioned problems do not share the same optimal solutions. The approximation ratio saturates to an $N$-dependent value compatible with $1-\alpha_\textrm{saturation}\sim N^{-1}$ up to $N=1,024$ (not enough iterations were performed to observe the saturation for larger values of $N$).

The number of iterations can be converted to a run-time following Appendix~\ref{app:iterations_to_run-time}. We present run-time data in the main text.

\begin{figure}[!t]
    \centering
    \includegraphics[width=1\columnwidth]{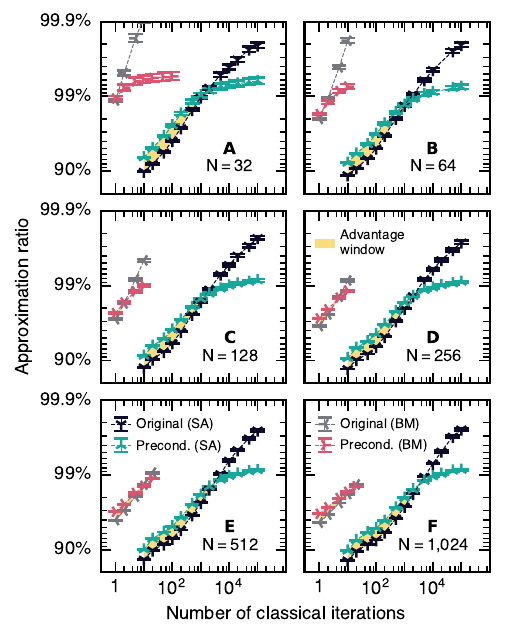}
    \caption{Average performance (approximation ratio versus the number of iterations) for $N$-variable Sherrington-Kirkpatrick spin glass problems via the classical SA and BM solvers based on the original and quantum-preconditioned $(p=1)$ problems. (A) $N=32$ variables. (B) $N=64$ variables. (C) $N=128$ variables. (D) $N=256$ variables. (E) $N=512$ variables. (F) $N=1,024$ variables. Each data point is averaged over $200$ randomly generated problem instances. Error bars indicate the standard error of the mean.}
    \label{fig:sk_N_preconditioning}
\end{figure}

\begin{figure*}[!t]
    \centering
    \includegraphics[width=1\textwidth]{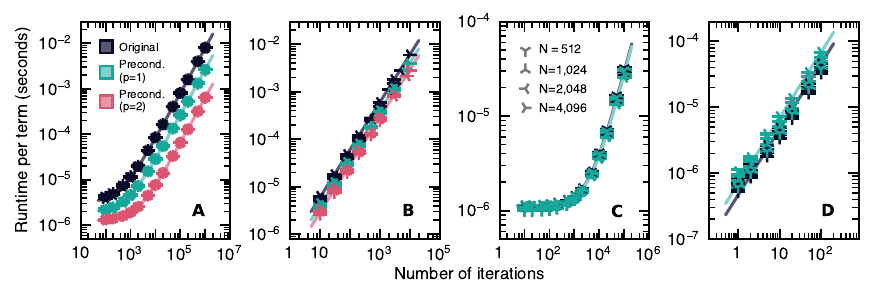}
    \caption{Average run-time per term in seconds (based on 64GB MacBook Pro with an Apple M1 Max chip) versus the number of iterations for original and preconditioned problems. (A) SA solver for random $3$-regular graph maximum-cut problems. (B) BM solver for random $3$-regular graph maximum-cut problems. (C) SA solver for Sherrington-Kirkpatrick spin glasses. (D) BM solver for Sherrington-Kirkpatrick spin glasses. Error bars indicate the standard error of the mean. Bold lines are fits according to the main text.}
    \label{fig:iterations_to_runtime}
\end{figure*}

\subsection{Sherrington-Kirkpatrick spin glasses}

Next, we provide additional data on Sherrington-Kirkpatrick spin glasses. We investigate the performance of SA and the BM solver on the original and quantum-preconditioned ($p=1$) problems. Problem sizes ranging from $N=32$ to $N=1,024$ are considered ($N=2,408$ was considered in the main text in Fig.~\ref{fig:sk_preconditioning}). We report data in Fig.~\ref{fig:sk_N_preconditioning}. We find that an advantage for the preconditioned problem for the BM solver only appears for large problem sizes. We highlight this advantage window in yellow. The average approximation ratio saturation value observed for a large number of iterations, $\alpha\simeq 99.3\%$, seems to be roughly independent of $N$.

\section{From Number of Iterations to Run-time}
\label{app:iterations_to_run-time}

Instead of using the run-time, the performance of the SA and BM solvers can be estimated through the number of iterations, which is a platform-independent number.

In both solvers, an iteration consists of $O(n)$ steps, where $n$ is the number of terms. For instance, in the case of SA, an iteration is also known as a sweep. For an $N$-variable problem, a sweep involves a local update on $O(N)$ variables, which is either accepted or refused based on the updated objective value of the proposed solution. Computing the updated objective value involves an arithmetic operation per term relating to the updated variable. Hence, the complexity cost per iteration is $O(n)$. A similar argument can be developed for the BM solver.

The run-time $t$ of a solver scales asymptotically with $N$ as
\begin{equation}
    t\bigr/n=An_\textrm{iter} + B,
    \label{eq:iter_to_run-time}
\end{equation}
where $n$ is the number of terms in the problem and $n_\textrm{iter}$ is the number of iterations. Here, $A$ and $B$ are solver-, platform-, and problem-dependent constants. In the case of SA, the constant $B$ is nonzero and corresponds to a one-time $O(n)$ heuristic for determining the temperature schedule based on the problem's terms (see Appendix~\ref{app:classical_solvers}). This is zero for the BM solver, which does not have any hyperparameters.

We show the relation between the run-time and number of iterations in Fig.~\ref{fig:iterations_to_runtime} for the random $3$-regular graph maximum-cut problems and Sherrington-Kirkpatrick spin glasses for SA and the BM solver on the original and preconditioned problems. The run-time is based on a 64GB MacBook Pro with an Apple M1 Max chip. In the case of $N$-variable $3$-regular graphs, we have $n=1.5N$ terms in the original problem, $n\simeq 4.5N$ at $p=1$, and $n\simeq 22.5N$ at $p=2$ in the large-$N$ limit (see Appendix~\ref{app:n_nonzero_terms_3reg}). For Sherrington-Kirkpatrick spin glasses, the number of terms is $n=N(N-1)/2$ in all cases. We reported the average approximation ratio as a function of both the number of iterations and run-time for the grid energy problem in Fig.~\ref{fig:mpes_preconditioning}.

\bibliography{bibliography}

\end{document}